\documentclass[aps,twocolumn,nofootinbib,superscriptaddress,amsfonts,floatfix]{revtex4-2} 
\usepackage{graphicx,amsmath,amssymb,amstext}
\usepackage{amssymb,amsbsy,amsfonts,amsthm,color,bm}
\usepackage{mathtools}
\usepackage{epsfig}
\usepackage{graphicx}
\usepackage{subfigure}
\usepackage{hyperref}
\usepackage{cancel}
\usepackage{balance}
\usepackage{booktabs}
\usepackage{diagbox,multirow}
\graphicspath{{Figures/}}
\usepackage[normalem]{ulem}

\usepackage{color}
\usepackage[dvipsnames]{xcolor}
\usepackage{soul}

\renewcommand{\eqref}[1]{Eq.~(\ref{#1})}
\newcommand{\eqsref}[1]{Eqs.~(\ref{#1})}
\newcommand{\Beq}{\begin{eqnarray}}
\newcommand{\Eeq}{\end{eqnarray}}

\def\lsim{\mathrel {\vcenter {\baselineskip 0pt \kern 0pt \hbox{$<$} \kern 0pt \hbox{$\sim$} }}}

\newcommand{\grchombo}{\mathtt{GRChombo}}

\newcommand{\dd}{\mathrm{d}}

\newcommand{\pd}{\partial}
\begin{document}

\title{Gravitational Magnus effect from scalar dark matter}

\author{Zipeng Wang}
\email{zwang264@jhu.edu}
\affiliation{Department of Physics and Astronomy, Johns Hopkins University, Baltimore, MD 21218, USA}

\author{Thomas Helfer}
\email{thomashelfer@live.de}
\affiliation{Institute for Advanced Computational Science, Stony Brook University, Stony Brook, NY 11794 USA}

\author{Dina Traykova}
\email{dina.traykova@aei.mpg.de}
\affiliation{Max Planck Institute for Gravitational Physics (Albert Einstein Institute), Am M\"uhlenberg 1, Potsdam-Golm, 14476, Germany}

\author{Katy Clough}
\email{k.clough@qmul.ac.uk}
\affiliation{School of Mathematical Sciences, Queen Mary University of London Mile End Road, London, E1 4NS, UK}

\author{Emanuele Berti}
\email{berti@jhu.edu}
\affiliation{Department of Physics and Astronomy, Johns Hopkins University, Baltimore, MD 21218, USA}


\begin{abstract}
In fluid dynamics, the Magnus effect is the force perpendicular to the motion of a spinning object as it moves through a medium. In general relativity, an analogous effect exists for a spinning compact object moving through matter, purely as a result of gravitational interactions. 
In this work we consider a Kerr black hole moving at relativistic velocities through scalar dark matter that is at rest. We simulate the system numerically and extract the total spin-curvature force on the black hole perpendicular to its motion.
We confirm that the force scales linearly with the dimensionless spin parameter $a/M$ of the black hole up to $a/M = 0.99$, and measure its dependence on the speed $v$ of the black hole in the range $0.1 \le v \le 0.55$ for a fixed spin. Compared to previous analytic work applicable at small $v$, higher-order corrections in the velocity are found to be important: the total force is nonzero, and the dependence is not linear in $v$.
We find that in all cases the total force is in the opposite direction to the hydrodynamical analogue, although at low speeds it appears to approach the expectation that the Weyl and Magnus components cancel. 
Spin-curvature effects may leave an imprint on gravitational wave signals from extreme mass-ratio inspirals, where the secondary black hole has a nonnegligible spin and moves in the presence of a dark matter cloud. We hope that our simulations can be used to support and extend the limits of analytic results, which are necessary to better quantify such effects in the relativistic regime. 
\end{abstract}


\maketitle


\section{Introduction} \label{sec:intro}


In classical fluid dynamics, the Magnus effect is the phenomenon of a spinning object, moving through a fluid, experiencing a force perpendicular to its motion and its spin~\cite{Morikawa,rubinow_keller_1961, munson_munson_2016}. 
An analogous effect is also present in the case of a black hole (BH) or other compact object due to gravitational interactions when it is subject to a current of matter or energy that is not aligned to its spin. Such an object will then experience a force orthogonal to both the spin and the matter current~\cite{Font:1998sc,Okawa:2014sxa,Cashen:2016neh,Costa:2018gva}. 

The Magnus effect in fluid dynamics is caused by contact interactions between the fluid and the spinning body at its boundary, which reduce the relative pressure experienced by the body on the side that is corotating with the flow. This leads to a net force perpendicular to the motion and the spin (in the corotating direction). The gravitational analogue, on the other hand, only involves gravitational interactions between the surrounding cloud and the compact object and is not a contact force. 
In the general relativistic (GR) picture, the curvature of the spacetime displaces the passing matter differently on the corotating/counterrotating sides of the BH, leading to an asymmetry in the flow pattern. Stable orbits on the corotating side can also pass closer to the body, whereas on the counter rotating side they will tend to be accreted at the same distance. A higher matter density on one side of the body will tend to gravitationally attract it, as happens in dynamical friction~\cite{Cashen:2016neh}, and preferential accretion on one side will also alter its trajectory~\cite{Okawa:2014sxa}. In summary, there is a complex interplay that exchanges momentum between the matter and the spacetime curvature of (and around) the BH, leading to motion perpendicular to the matter flux and the spin. 

Whilst previous studies agreed that such a force should exist, they found conflicting results about the direction of the effect. References~\cite{Okawa:2014sxa, Cashen:2016neh} suggested that the gravitational effects produce an ``anti-Magnus'' force
\footnote{The usual Magnus force has a direction given by $\sim {\bf v} \times {\bf \omega}$, where ${\bf v}$ is the velocity of the fluid relative to the body and ${\bf \omega}$ the spin direction of the body, whereas these authors suggested results of the form $\sim {\bf \omega}\times {\rm v}$. In more intuitive language, the authors found a force towards the side of the BH that is counterrotating with respect to the matter flux.}. However, Ref.~\cite{Font:1998sc} argued for a gravitational Magnus force in the same direction as the hydrodynamical Magnus force when considering nonaxisymmetric relativistic Bondi-Hoyle accretion. 

These differences were resolved in Ref.~\cite{Costa:2018gva}, where the force was studied in a more rigorous post-Newtonian analysis. This work highlighted that the total spin-curvature force (which the previous works had tried to measure) is composed not just of a gravitational Magnus force, which has a well-defined form and a direction that is the same as the hydrodynamical case, but also a ``Weyl force'', coming from the magnetic part of the Weyl tensor, which is highly dependent on the physical scenario and boundary conditions. Physically speaking, we are interested in the total force, composed of both Magnus and Weyl, since it is this that determines the overall motion of the BH in the given scenario. The direction of the total force is scenario-dependent and in the case we study in this paper we find it to be overall ``anti-Magnus'', in agreement with the work in Refs.~\cite{Okawa:2014sxa, Cashen:2016neh}, since it is dominated by the dynamical friction-like effect caused by the enhanced density of the flow on the counterrotating side (which one can identify with the Weyl component).

Such spin-curvature effects could produce a detectable change in gravitational wave signals where the progenitors are located in a dense matter environment.
Extreme mass-ratio inspirals consist of a massive BH with mass $\sim 10^6 M_{\odot}$ such as those observed at the center of galaxies, and a secondary BH with mass comparable to $M_{\odot}$~\cite{Amaro-Seoane:2012lgq,LISAConsortiumWaveformWorkingGroup:2023arg}. During the long inspiral of the secondary~\cite{Berry:2019wgg}, the presence of matter around the massive BH can affect its trajectory through effects such as  modifications of the mass distribution~\cite{Hannuksela:2018izj}, dynamical friction~\cite{Traykova:2021dua, Traykova:2023qyv,Vicente:2022ivh}, and Magnus-like effects~\cite{Costa:2018gva}, potentially leaving an imprint on the gravitational wave signal detectable by LISA~\cite{Barausse:2014tra,Bonga:2019ycj}. In Ref.~\cite{Costa:2018gva} it was shown that the gravitational Magnus effect causes orbital precession in extreme mass-ratio inspirals with a unique modulating effect on the gravitational wave signal that differentiates it from other causes. 

For the case of dark matter (DM) environments, on which we focus in this work, the energy densities required to have significant effects on the gravitational wave signal are high relative to the expected average galactic values~\cite{Nesti:2013uwa,Barausse:2014tra,Pato:2015dua,DeMartino:2018zkx,Li:2020qva,Ablimit:2020gxw}. 
However, average galactic densities describe DM on large scales only, and the distribution on smaller scales around BHs is less well constrained. Several mechanisms exist that could create DM overdensities around isolated BHs. One is the superradiant instability, in which a bosonic field extracts energy and angular momentum from a highly spinning BH via repeated scattering in the ergoregion~\cite{1971JETPL..14..180Z,Press:1972zz,Zouros:1979iw,Detweiler:1980uk,Cardoso:2004nk,Cardoso:2005vk,Dolan:2007mj,Arvanitaki:2010sy,Arvanitaki:2014wva,Herdeiro:2014goa,East:2017ovw} (see~\cite{Brito:2015oca} for a review).
Another is accretion in the potential well around BHs, which creates ``dark matter spikes''~\cite{Gondolo:1999ef,Sadeghian:2013laa,Ferrer:2017xwm,Speeney:2022ryg,Speeney:2024mas}. 
These spikes have a power law profile with an exponent that depends on the effective equation of state of the DM~\cite{Gnedin:2003rj,Merritt:2003qk,Merritt:2006mt,Sadeghian:2013laa,Shapiro:2022prq,DeLuca:2023laa,Berezhiani:2023vlo}. For low-mass, wavelike DM candidates, their form is dependent on the relative Compton wavelength of the DM particle and the BH horizon~\cite{Sanchis-Gual:2016jst,Clough:2019jpm,Hui:2019aqm,Bamber:2020bpu,Bucciotti:2023bvw,deCesare:2023rmg}. Such overdensities have recently been shown to persist even through equal-mass mergers~\cite{Bamber:2022pbs,Aurrekoetxea:2023jwk}.

\begin{figure}[h]
  \includegraphics[width=0.45\textwidth]{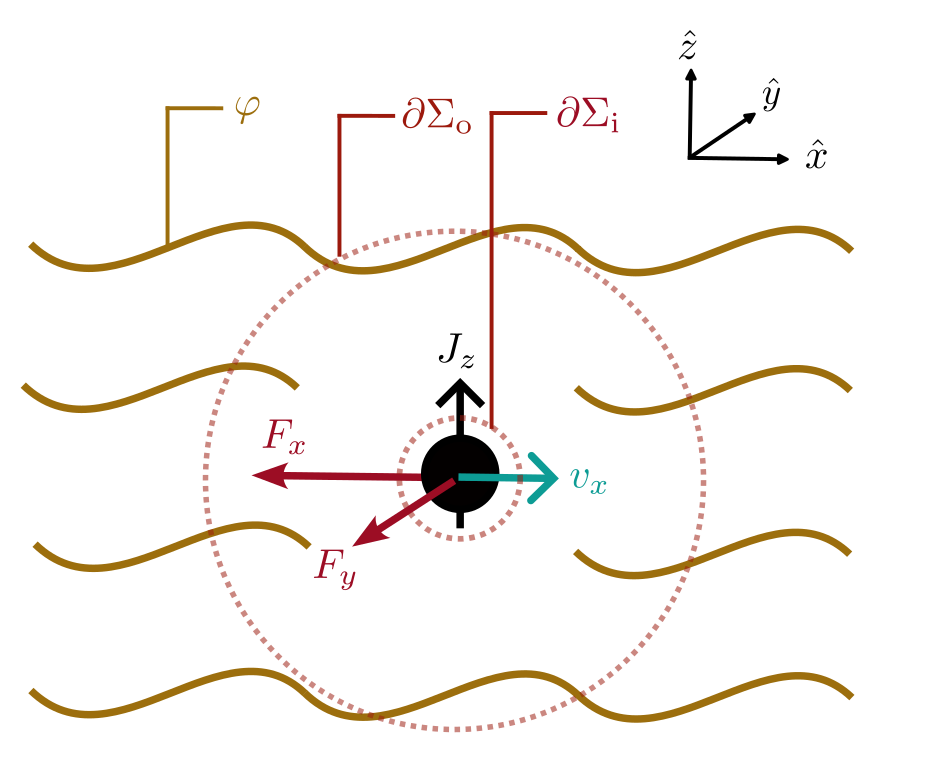}
  \caption{A schematic illustration of the setup of the numerical simulations used to extract the total spin-curvature force. 
  A boosted Kerr BH is located at the center of the computational domain, with its velocity pointing in the $+x$ direction and its angular momentum pointing in the $+z$ direction. The distribution of the scalar field $\varphi$ changes as a result of the curvature of the metric background, resulting in effective forces on the BH, $F_x$ (dynamical friction) and $F_y$ (spin-curvature). To quantify these forces, we integrate quantities related to the scalar field momentum within the spherical boundaries $\partial \Sigma_o$ and $\partial \Sigma_i$ and fluxes of the scalar field momentum on $\partial \Sigma_i$. Further details are given in the main text.
  }
  \label{fig:cartoon}
\end{figure}

In this paper, we investigate numerically the magnitude of the total effective spin-curvature force on a Kerr BH. We model a scalar field surrounding the BH with an asymptotically homogeneous density, as a model for a light bosonic DM environment (e.g., axions~\cite{Marsh:2015xka,Peccei:2006as}). Our particular setup for measuring the force is illustrated in Fig.~\ref{fig:cartoon}, where the central black circle represents the Kerr BH, the $J_z$ arrow represents the direction of its angular momentum, and the $v_x$ arrow labels the direction of its velocity. Due to the presence of the scalar field $\varphi$, the BH experiences a force in the $-x$ direction, which has been studied in a similar setup in previous work~\cite{Traykova:2021dua, Traykova:2023qyv}. The force of interest for this work is the one in the $-y$ direction, perpendicular to the direction of motion\footnote{As pointed out in~\cite{Costa:2018gva}, it should properly be thought of as composed of both the Magnus force and the Weyl force, and referred to as the total spin-curvature force.}
. These forces are labeled with the red arrows in Fig.~\ref{fig:cartoon}. As in the previous work, we simulate the scalar field on a fixed metric background using the \texttt{GRDzhadhza} code~\cite{Aurrekoetxea:2023fhl}, and as such our results are valid only to first order in the density of the matter. We extract the total force on the BH in the $y$ direction by tracking the momentum fluxes of the simulated scalar field. We fix the scalar field mass as $\mu = 0.2 M^{-1}$, and study the effect of varying the spin and the velocity of the Kerr BH on the resultant force.

The rest of this paper is organized as follows. In Sec.~\ref{sec:setup}, we provide the scalar field evolution equations, the background metric on which it evolves, the key theoretical background to which we compare our results, and details of our numerical setup and method to extract the spin-curvature force on the BH. In Sec.~\ref{sec:results}, we present and discuss the magnitude of the force extracted from our simulations and its dependence on the BH spin and velocity. 
We summarize our results and discuss future directions in Sec.~\ref{sec:discussion}. In Appendix~\ref{app:validation} we provide further details of the numerical tests we undertook to validate our results. Throughout the paper we use geometrical units ($G=c=1$).

\section{Setup and numerical methods}
\label{sec:setup}
\subsection{Scalar field}
We consider a complex scalar field, $\varphi$, minimally coupled to gravity. The dynamics of such field are described by the following action,
\begin{equation}
    \label{eqn: action}
    S=\frac{1}{2}\int \dd^4x \sqrt{-g}\left( \nabla_\mu \varphi^* \nabla^\mu \varphi- \mu^2 |\varphi|^2 \right)\,,
\end{equation}
 where $g$ is the determinant of the spacetime metric and $\mu$ is the mass of the scalar field. Varying the field $\varphi$ with this action results in the Klein-Gordon equation
\begin{equation}
\label{eqn:4d_KG}
    \left(\Box_g-\mu^2\right) \varphi=0\,.
\end{equation}

Solving \eqref{eqn:4d_KG} requires information on the spacetime metric, which in the 3+1 ADM decomposition has the general form,
\begin{equation}
\label{eqn: ADM}
\dd s^2=-\alpha^2\dd t^2+\gamma_{ij}\left(\dd x^i + \beta^i\dd t\right)\left(\dd x^j + \beta^j\dd t\right)\,,
\end{equation}
where $\alpha$ is the lapse function, $\beta_i$ is the shift, and $\gamma_{ij}$ is the spatial metric. The unit vector normal to the spatial slices is $n^\mu = \left(1/\alpha, -\beta^i/\alpha \right)$. The extrinsic curvature is defined as
\begin{equation}
    K_{ij} =\alpha^{-1} D_{(i} \beta_{j)} \,,
\label{eqn:extr_curv}
\end{equation}
where $D_i$ is the covariant derivative compatible with the spatial metric $\gamma_{ij}$. 
We can then recast \eqref{eqn:4d_KG} as evolution equations for the real part and the imaginary part of the scalar field. Explicitly, these take the form
\begin{align}
\partial_t \varphi &= \alpha \Pi +\beta^i\partial_i \varphi \label{eqn:dtphi} ~ , \\ 
\partial_t \Pi &= \alpha \gamma^{ij}\partial_i\partial_j \varphi +\alpha\left(K\Pi -\gamma^{ij}\Gamma^k_{ij}\partial_k \varphi - \mu^2 \varphi\right)\nonumber  \\
& + \partial_i \varphi \partial^i \alpha + \beta^i\partial_i \Pi \label{eqn:dtPi} ~ ,
\end{align}
where $\Pi$ is the conjugate momentum of the real and imaginary parts of the scalar field $\varphi$, and $\Gamma_{ij}^k$ is the Christoffel symbol associated with $\gamma_{ij}$. In this work the metric functions $\alpha$, $\beta^i$ and $\gamma_{ij}$ are related to a boosted Kerr background, and their specific form is given in the following section.

We set the initial conditions so that the scalar field is uniform in the simulation box, with $\varphi(t=0) = \varphi_0$, and $\Pi(t=0) = i\mu \varphi_0 $. Therefore, the asymptotic density of the scalar field in its rest frame is
\begin{equation}
    \rho = \frac{1}{2} \mu^2 \varphi_0^2 + \frac{1}{2} \left\rvert\Pi_0^2 \right\rvert = 
    \mu^2 \varphi_0^2.
\end{equation}
Since we do not account for the backreaction of the scalar onto the metric, the amplitude of the scalar field can be arbitrarily chosen: the result simply rescales with the physical value of $\rho$ (assuming $\rho M^2 \ll 1$). In all simulations we set $M=1$, $\mu M=0.2$, and $\varphi_0 = 0.1$ in the geometrical units ($G=c=1$) used in the code.

\subsection{Fixed metric background and choice of frame}
\label{sec:background}

Following Re.~\cite{Okawa:2014nda}, we write the Kerr metric in the quasi-isotropic coordinates $(\bar t,  \bar r, \bar \theta, \bar \phi)$ as
\begin{align}
 \dd s^2 = - \left( 1- \frac{2 M r_{\mathrm{BL}}}{\Sigma} \right) \dd \bar t^2  \nonumber 
 &+ \psi_0^4 \Bigg[ \frac{\left(\bar r + \frac{r_+}{4}\right)^2}{\bar r \left( r_\mathrm{BL} - r_- \right)} \dd \bar r^2  \nonumber \\
 + \bar r^2 \dd \theta^2 &+ \frac{\mathcal{A}}{\Sigma^2} \bar r^2 \sin^2 \bar \theta \dd \bar \phi^2  \Bigg] \medspace \label{eqn: metric}, 
\end{align}
where we define
\begin{align}
    \mathcal{A} &= \left( r_{\mathrm{BL}}^2 + a^2 \right)^2 - \Delta a^2 \sin^2 \bar \theta \medspace , \\
    \Sigma &= r_{\mathrm{BL}}^2 + a^2 \cos^2 \bar \theta \medspace , \\
    \Delta &= r_\mathrm{BL}^2 - 2 M r_\mathrm{BL} + a^2 \medspace .
\end{align}
Here, $M$ is the mass of the BH, $a$ is the spin parameter, $r_{\mathrm{BL}} = \bar r \left( 1 + \frac{r_+}{4\bar r} \right)^2$ is the Boyer-Lindquist radius, and $r_\pm = M \pm \sqrt{M^2 - a^2}$ is the location of the horizons in Boyer-Lindquist coordinates.

As illustrated in Fig.~\ref{fig:cartoon}, we need to create a relative motion between the Kerr BH and the simulated scalar field. To achieve this, we boost Eq.~(\ref{eqn: metric}) in the $x$ direction. However, naively applied, this boost would result in the BH moving across the simulation grid over time. To remedy this, we also apply a Galilean transformation ($x \rightarrow x - v t$) so that the the computational grid tracks the movement of the BH. 
The result is that the normal observers are in the rest frame of the scalar field, and observe a boosted, length-contracted BH, that moves in the positive $x$ direction. The time-like observers, on the other hand, move with the BH, and so for them the BH does not move (although note that they are {\it not} simply in the rest frame of the BH, as they observe it to be length-contracted and the matter to be at rest).
The resulting transformation is, as in Ref.~\cite{Traykova:2021dua},
\begin{equation}
\label{eqn: transform}
\bar t = t/\gamma - \gamma v x \qquad \bar x = \gamma x \qquad \bar y = y \qquad \bar z = z,  
\end{equation}
where $v$ is the velocity of the BH and $\gamma$ the Lorentz boost.
We call the resulting frame $(t,  x, y,  z)$ the ``simulation frame''. The original barred coordinates $(\bar t,  \bar x, \bar y,  \bar z)$ are the BH rest frame, in which it is most convenient to express our results to compare with the existing literature.

The lapse $\alpha$, shift $\beta_i$, the spatial metric $\gamma_{ij}$, and the extrinsic curvature $K_{ij}$ can be obtained by applying \eqref{eqn: transform} to \eqref{eqn: metric}, and comparing the result against \eqref{eqn: ADM}. In the BH rest frame, $(\bar t, \bar r, \bar \theta, \bar \phi)$, these quantities can be easily read off from the metric, but in the simulation frame $(t, r, \theta, \phi)$, 
the analytical expressions are complicated (although straightforward to implement numerically). See Sec.~\ref{sec:numerics} for more details on the numerical implementation of the metric background.

\begin{figure*}[ht]
  \includegraphics[width=1\textwidth]{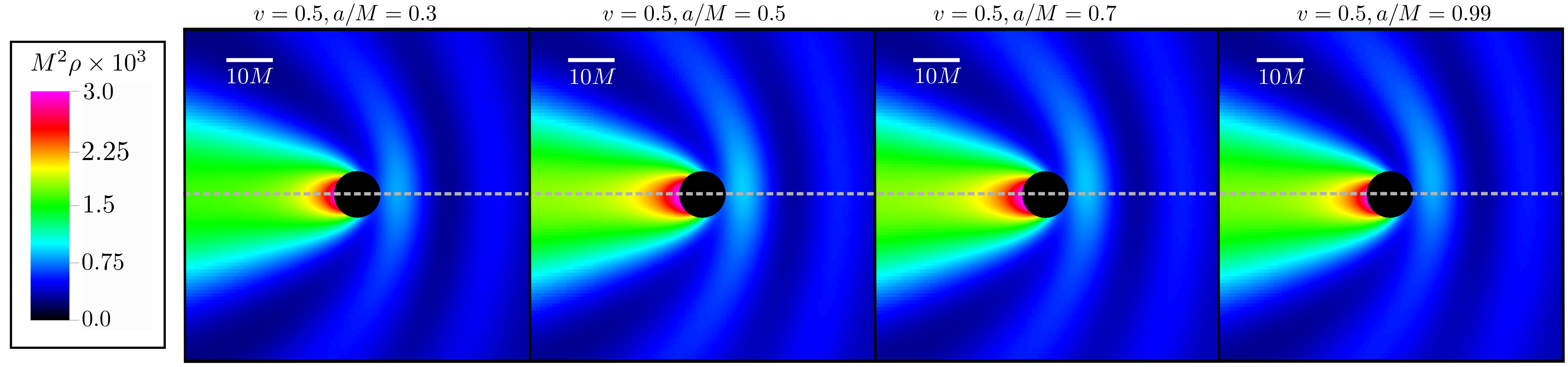}
  \caption{The density $\rho$ of the scalar field surrounding the BH at late times of the simulations with $v=0.5$ and different BH spins $a/M$. The central black circle shows the inner extraction boundary $r_{\rm i}/M = 5$. The grey dashed line shows the location of the $x$ axis. The angular momentum of the BH points to the $z$ axis (out the page). We see that as the spin increases the matter density is rotated so that it is higher in the $-y$ direction, resulting in an attractive force in this direction that dominates the total spin-curvature force. Here, we choose late-time snapshots of $\rho$ where the field is in a roughly steady state, although small oscillations in time still persist, as can be seen in Fig.~\ref{fig:force components}.
  }
  \label{fig:a_dependence_screenshot}
\end{figure*}

\subsection{Quantification of the gravitational spin-curvature force}

We follow the approach described in Refs.~\cite{Clough:2021qlv, Traykova:2021dua,Traykova:2023qyv} to quantify the effective force on the BH in the direction perpendicular to its motion. Strictly speaking, this is not a gravitational force, since we are working in a general relativistic framework, but rather it is an exchange of momentum between the matter and the curvature. However, given that we can identify observers in the asymptotically flat region, and that we reach a stationary state of the matter/BH system, we can define a force 4-vector on the BH $F^\mu \equiv \frac{\partial P^\mu}{\partial \tau}$, where $P^\mu$ is the momentum of the spacetime as measured by the asymptotic observer and $\tau$ is their proper time. We note that this approach relies on an action-reaction principle in assuming that the force on the matter is the inverse of the force on the BH. We are therefore, as pointed out in Ref.~\cite{Costa:2018gva}, measuring the total spin-curvature force, and not simply the Magnus force. This will be discussed further in Sec.~\ref{sec:theory} below.

Our simulations neglect the backreaction of the matter onto the spacetime, which means that our measured force is valid up to first order in $\rho M^2$ (the density as compared to the curvature of the BH). This would be a reasonable approximation for most physical scenarios involving DM: even in the best cases for superradiant vector clouds, this number is of order $10^{-5}$~\cite{Brito:2014wla,East:2018glu}. There is also a strong advantage to working in such a fixed background, compared to full NR as in~\cite{Okawa:2014sxa}, in that there is a well defined gauge in which to interpret the results.

\begin{figure*}[t]
  \includegraphics[width=0.96\textwidth]{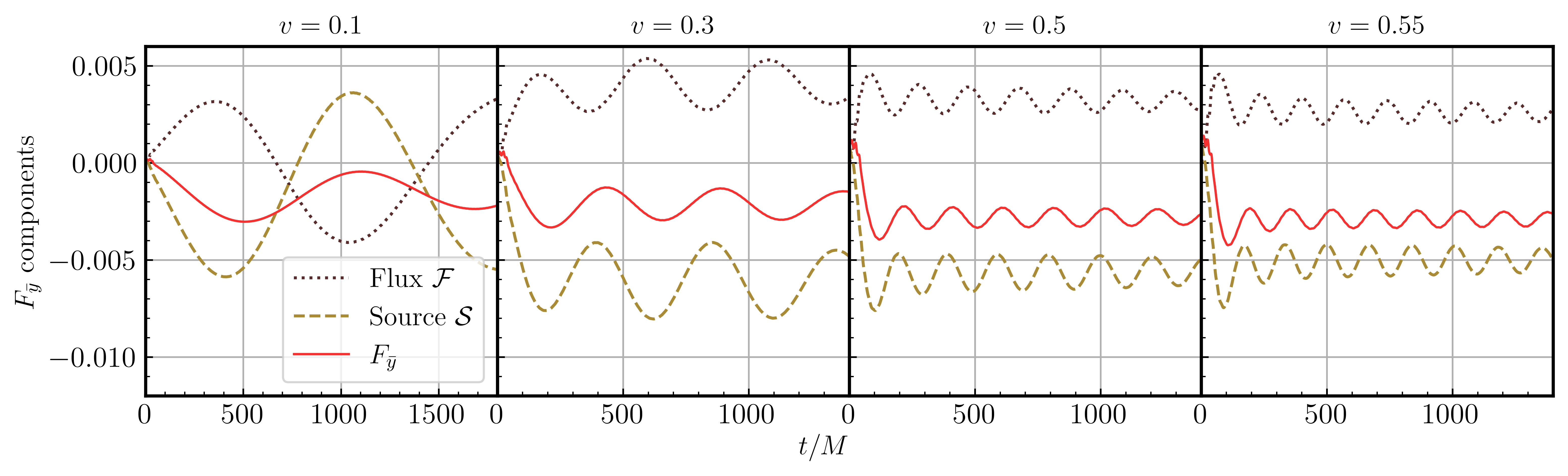}
  \caption{
In this plot we separate out the contributions to the overall force from the flux $\mathcal F$ and the source term $\mathcal S$, which as discussed in the main text can be loosely identified with the Magnus and the Weyl parts of the spin-curvature force. Each panel represents a different speed $v$, giving rise to the points in third panel of Fig.~\ref{fig:trends}. We use the BH spin parameter $a/M=0.7$ in all simulations shown here. We see that at smaller speeds the contributions are roughly equal and opposite, which is consistent with Ref.~\cite{Costa:2018gva}. As $v$ increases the source term dominates, and we obtain an overall negative force in the $y$ direction.
  }
  \label{fig:force components}  
\end{figure*}

The energy-momentum tensor of the scalar field, $\varphi$, with the action in \eqref{eqn: action} has the form
\begin{equation}
    T_{\mu \nu}=\nabla_{(\mu}\varphi^* \nabla_{\nu)}\varphi-\tfrac{1}{2}g_{\mu \nu}\left[\nabla_\delta \varphi^* \nabla^\delta \varphi +\mu^2|\varphi|^2\right]\,.
\end{equation}
We define its projections normal to and into the three-dimensional spatial slice of the spacetime as,
\begin{equation}\label{eqn:rhoSdecomp}
    \rho \equiv n_{\alpha} n_{\beta} T^{\alpha\beta},
    \quad\!\!\!\!
    S_i \equiv -\gamma_{i\alpha}n_{\beta} T^{\alpha\beta},
    \quad\!\!\!\!
    S_{ij} \equiv\! \gamma_{i\alpha}\gamma_{j\beta} T^{\alpha\beta}.
\end{equation}
Following Refs.~\cite{Clough:2021qlv, Traykova:2021dua,Traykova:2023qyv}, we extract the backreaction force in the $y$ direction by the scalar field on the BH in the simulation frame as
\begin{multline}
F_y = \partial_t P_y
= \underbrace{-\int_{\partial{\Sigma_{\rm i}}} \dd A_j\, \alpha T^j_{y}}_{{\rm ``flux"} \mathcal{F}} \\   
\underbrace{-\int_{{\Sigma_{\rm o}} - \Sigma_{\rm i}}\dd^3x \sqrt{-g}\, T^\mu_\nu \,^{(4)}\Gamma^\nu_{\mu y}}_{{\rm ``source"} \mathcal{S}}  \,,
\label{eqn:force-extr}
\end{multline}
where~$\dd A_j\coloneqq \dd^2x\sqrt{\sigma} N_j$, where~$\sigma$ is the determinant of the induced metric on the 2-surfaces~$\pd \Sigma$, and~$N_j$ is their outward pointing normal, normalized such that $\gamma_{ij} N^i N^j = 1$. Here $^{(4)}\Gamma^\nu_{\mu i}$ is the Christoffel symbol of the full four-dimensional metric (see~\cite{Clough:2021qlv} for its expression in 3+1 decomposed variables, and further implementation details), and $\Sigma_{\rm i} \subset \Sigma_{\rm o}$ are three-dimensional spherical volumes centered around the singularity (see Fig.~\ref{fig:cartoon} for an illustration). 

The first term in \eqref{eqn:force-extr}, which we refer to as the flux $\mathcal{F}$, is the change in the momentum of the BH due to accretion of the scalar field into the region $\Sigma_{\rm i}$ near the BH. The second term, which we refer to as the source $\mathcal{S}$, is the momentum exchange due to gravitational interactions between the matter and the curvature (we see explicitly that it is due to the coupling between the Christoffel symbol and the stress-energy tensor in the surrounding volume). In the Newtonian limit the second term is simply the gravitational attraction on the BH due to the integrated effect of the unevenly distributed surrounding matter (see Fig.~\ref{fig:a_dependence_screenshot}). The component in the $x$ direction is the dynamical friction, and the component in the $y$ direction arises due to the spin-curvature effects. 
The total force $F_i$ is a well-defined quantity, while its split between the two terms on the right-hand-side of \eqref{eqn:force-extr} is dependent on the slicing and on the exact choice of $\Sigma_{\rm i}$. Nevertheless, we can loosely identify the former part with the Magnus part of the force and the latter with the Weyl part of the force (see Sec.~\ref{sec:theory} below, and Appendix~\ref{app:correspondence} for further details and a justification in the Newtonian limit). 

To obtain the correct 4-vector component of the force, we need to account for the difference between the asymptotic proper time (also the time measured by observers in the BH rest frame) compared to our simulation frame. We therefore find the $\bar{y}$ component of the covariant spin-curvature force in the BH rest frame to be
\begin{equation}
    \begin{gathered}
        F_{\bar y} = \partial_{\bar{t}} P_y = \gamma F_{y}\,,
    \end{gathered}
    \label{eqnF_transform}
\end{equation}
using the coordinate transformations in \eqref{eqn: transform}.

The contributions to the extracted force, as shown in Fig.~\ref{fig:force components}, are found to be oscillatory in time. This appears to be due to our (somewhat artificial) choice to start evolving the scalar field from a uniform distribution. We observe that the total force evolves to a steady average state, but a long-lived oscillation around that state is present. 
We treat the amplitude of these oscillations as the uncertainty on the force we measure, as marked by the error bars in Fig.~\ref{fig:trends}. The amplitude appears to gradually decrease in time, so in principle we could extend the simulation to later times to further constrain the force. However, the damping time of these oscillations is long, so reducing them requires significantly longer simulations, which are computationally infeasible. 
In particular, numerical errors and boundary effects accumulate over time, and these affect the accuracy of the simulation (this is discussed further in Appendix~\ref{app:validation}).

\subsection{Numerical implementation}
\label{sec:numerics}

We use an adapted version of \texttt{GRDzhadzha}, an analytic background code~\cite{Aurrekoetxea:2023fhl} for efficiently evolving BH environments, based on the numerical relativity code $\grchombo$~\cite{Clough:2015sqa,Andrade:2021rbd}. This code evolves the scalar field equations,~\eqsref{eqn:dtphi} and (\ref{eqn:dtPi}), on the fixed Kerr background described above. We extract the induced force by evaluating \eqref{eqn:force-extr} in a fixed coordinate volume. The specific numerical setup is detailed below.

In all of our simulations, we set the computational domain size to be $L=1500$M in all directions, but taking advantage of the reflective symmetry in the $z$ direction to simulate only half of the domain. The grid spacing on the coarsest level is $\Delta_0 = 11.718$ and we have eight 2:1 refinement levels, centered at the BH. This gives the finest resolution $\Delta_8 = 0.046$ around the horizon of the BH. 

To describe the Kerr BH in the simulation coordinates $(t, r, \theta, \phi)$, we transform the metric \eqref{eqn: metric} (written in the BH coordinates $(\bar t,  \bar r, \bar \theta, \bar \phi)$) to the simulation coordinates according to \eqref{eqn: transform}. The expressions for the metric and its derivatives can be written down explicitly, but are extremely long, and we have found that their evaluation causes a significant slow-down of the simulations to around $6M$ per hour. To remedy this, we numerically store the values of the metric in the memory and evaluate the derivatives using finite differences. Using the resolution setup discussed above, the difference in the evolution of the scalar is negligible compared to using the exact derivatives, but significantly faster.
A different coordinate system, such as a boosted Kerr-Schild metric, would be more efficient, but would require a mapping of the observers (who are infalling) to the correct inertial asymptotic observers in order to obtain the correct force. This would require significantly rewriting the code, but it is an update we plan to implement in future work.

All results presented in this study are obtained with a scalar field mass parameter $\mu M = 0.2$. Higher masses require significantly higher spatial and time resolutions, so this is chosen for numerical convenience. As in the case of dynamical friction~\cite{Traykova:2021dua,Traykova:2023qyv, Vicente:2022ivh}, the scalar mass should have an impact on the forces, but we do not study that dependence in this work.

For the volume over which the force is evaluated, we put the inner and outer boundaries $\partial \Sigma_i$ and $\partial \Sigma_o$ at coordinate radii $r_i=5M$ and $r_o=700M$ in code units.
The outer boundary is sufficiently far out to ensure that we capture the asymmetric part of the cloud profile (this depends on the de Broglie wavelength of the scalar). It should also be reasonably far away from the computational boundary to reduce boundary effects. The inner boundary allows us to account for the region close to the horizon where we do not resolve the full dynamics, so it needs to be slightly outside that. It is also chosen so that there is no overlap between its surface and the mesh refinement boundaries, which reduces numerical noise. Unlike the dynamical friction force, which scales with $r$, the spin curvature force we measure seems to be independent of the precise choices of $r_i$ and $r_o$, as illustrated in the tests presented in Appendix~\ref{app:validation}. Physically, this is because the asymmetric configuration of the matter cloud around the BH in the y direction is concentrated relatively close to it, and so (beyond some radius) there is no further contribution to the force.


\subsection{Theoretical expectation and interpretation}
\label{sec:theory}

In Ref.~\cite{Costa:2018gva} the authors calculate the total spin-curvature force on a spinning particle moving though a DM cloud using a first-order post-Newtonian (1PN) approximation. They break the general formula for the total spin-curvature force in two parts, as
\begin{equation}
\label{eqn: PNformula}
    F^\mu_{tot} = F^\mu_{Mag} + F^\mu_{Weyl}.
\end{equation}
Here the Magnus force (having the same sign as its hydrodynamical equivalent) is
\begin{equation}
\label{eqn: PNformula2}
    F^\mu_{Mag} = 4 \pi \epsilon^\mu_{\nu\rho\sigma} J^\nu S^\rho U^\sigma,
\end{equation}
where $J^\mu = - T^\mu_\nu U^\nu$ is the energy current of the DM cloud measured by an observer moving with 4-velocity $U^\mu$, and $S^\mu$ is the spin 4-vector of the BH. In the case of a particle, this part is entirely determined by the local value of the stress energy tensor (and equivalently, by the Ricci tensor), and thus independent of the asymptotic behavior and boundary conditions. In our framework, we can roughly identify it with the flux term $\mathcal{F}$ defined above, although the precise value is, as noted already, highly dependent on the choice of $\partial\Sigma_i$.

The Weyl component is the projection of the spin vector onto the magnetic part of the Weyl curvature tensor
\begin{equation}
\label{eqn: FWeyl}
    F^\mu_{Weyl} = - B^{\mu\nu} S_\nu, \quad B_{\mu\nu} = \star C_{\mu\alpha\nu\beta} U^\alpha U^\beta ~.
\end{equation}
Since the Weyl curvature tensor conveys information on nonlocal matter inhomogeneities, governing tidal forces and gravitational radiation effects, it is very much determined by the physical scenario of interest, in particular the surrounding matter configuration and the boundary conditions. In our framework, it again makes sense to identify this contribution with the term that is not local to the BH horizon, that is, the source $\mathcal{S}$.

Our physical scenario in this paper is most similar to the slab geometry case described in Ref.~\cite{Costa:2018gva}, in particular to their ``case 2'', which is finite in the $z$ direction\footnote{It is not obvious to us why our results should differ from the slab geometry that is finite only in the $y$ direction, provided the finite size $h$ is larger than the size of the region in which the stationary end state of the cloud becomes asymmetric. The authors of Ref.~\cite{Costa:2018gva} discuss this inconsistency as a limit of the post-Newtonian approach, so this is perhaps resolved by the fully general relativistic treatment used here. In such a case they find that the Weyl force should be equal and in the same direction to the Magnus force. This is clearly inconsistent with our results, for which the contributions appear to be of opposite sign, as we will discuss in Sec.~\ref{sec:results}.}.
The expectation in this case is that the Weyl and Magnus parts of the force should be equal and opposite, and thus cancel out to zero. We see evidence that this state is approached in the low-$v$ limit, as shown in Fig.~\ref{fig:force components}, but at higher $v$ the Weyl part appears to dominate, such that the overall force is in the ``anti-Magnus'' direction.

In order to put our results into context, we compare our total force to the magnitude of the Magnus part of the force expected from Ref.~\cite{Costa:2018gva} (but including the factors of the boost $\gamma$ that naturally appear). Consider the rest frame of the BH, in which it has mass $M$, angular momentum per unit mass of $a$ (so $S^\mu = (0,0,0,aM)$), and 4-velocity $U^\mu = (1,0,0,0)$,  whilst the fluid 4-velocity is $V^\mu = (\gamma,-\gamma v,0,0)$ and so $J^\mu = T^\mu_0 = \rho V^\mu V_0 = (\rho \gamma^2, -\rho \gamma^2 v,0,0)$, with $\rho$ the energy density of the matter in its rest frame.
The resulting formula for the force $F_{Mag, PN}$ is
\begin{equation}
    F_{Mag, PN} = 4 \pi v \gamma^2 \rho a M ~,
\label{eq:MagPN}
\end{equation}
in the positive $y$ direction, where $\rho = \mu^2 \varphi^2_0$ in our simulation setup. In Fig.~\ref{fig:trends} we plot $|F_{Mag, PN}|$ against our simulation results to give an idea of the deviation from the expected values at relativistic speeds. 


\section{Results}
\label{sec:results}

\begin{figure*}[t]
  \includegraphics[width=0.98\textwidth]{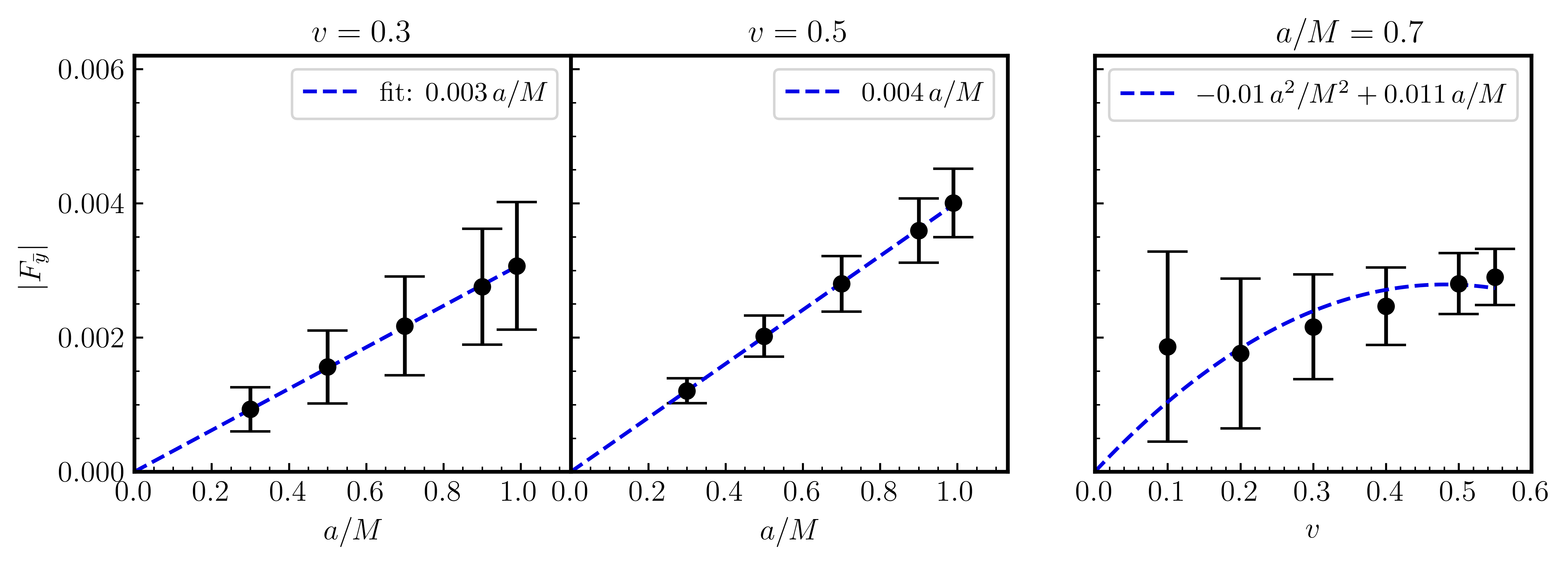}
  \caption{Magnitude of the total spin-curvature force induced by the scalar field with different BH spin parameters $a/M$ and velocities $v$. The left two panels show the extracted $y$ component of the force with different BH spins $a$ but the same BH velocity $v$ ($v=0.3$ and $v=0.5$, respectively). The right panel shows the Magnus force at fixed BH spin $a/M=0.7$ but different BH velocities . The error bars here show the amplitude of the oscillations between $t/M=1000$ and $t/M=1500$. The blue dashed line shows a fit to our results. As discussed in Sec.~\ref{sec:theory}, the total force would be expected to be zero for small $v$, but at relativistic speeds it shows an overall nonzero value in the ``anti-Magnus'' direction, in which the Weyl contribution dominates. 
  \label{fig:trends}}
\end{figure*}

Our numerical results are summarized in Fig.~\ref{fig:trends}. In the left two panels we observe a linear relationship between the magnitude of the total spin-curvature force $F_{\bar{y}}$ and the BH spin parameter $a$, at two different BH velocities, $v=0.3$ and $v=0.5$. This relationship between the angular momentum of the compact spinning object and the magnitude of the spin-curvature forces is consistent with the scaling in \eqref{eq:MagPN}. However, in the 1PN analysis of Ref.~\cite{Costa:2018gva}, the Magnus and Weyl parts of the force cancel exactly, and so one expects zero net force. For these larger values of $v$, it is clear that the behavior changes, with the Weyl part of the force dominating, such that the net force is in the ``anti-Magnus'' direction, the direction where the flow is counterrotating with the BH spin. 

The nontrivial dependence on $v$ is shown in the third panel of Fig.~\ref{fig:trends}, where we now plot the magnitude of the force against the velocity at one BH spin, $a/M=0.7$. 
We have fitted a quadratic in $v$ to the results simply to provide a numerical fit, but further work is needed to understand the functional form. We can see that the simple addition of appropriate boost factors to the 1PN Magnus force to account for the increase in the relative flux, as in \eqref{eq:MagPN}, does not appear to account well for the results.

Whilst we do not present the results here, for $v>0.6$ we see tentative evidence for a drop in the magnitude of the Magnus force, and even hints that it may change sign, which supports the use of a quadratic fit. However, it is hard to simulate BHs with velocities higher than $v=0.55$ using our current numerical setup. At higher velocities, the scalar field wavelength becomes smaller in the simulation coordinates, making it harder to resolve the field. Our grid refinement scheme can resolve fine features of the scalar field in the vicinity of the BH. However, the higher-velocity simulations require higher resolutions across the whole simulation domain, which is prohibitively expensive in our numerical setup. We suggest ways to address this limitation in Sec.~\ref{sec:discussion}.

We also track the dynamical friction force in the $x$ direction, and find that it is not changed significantly compared to previous results~\cite{Traykova:2021dua,Traykova:2023qyv}. The spin-curvature force is approximately 1\% of the dynamical friction force, and as such any effect it has on the latter is undetectable within the large error bars of that measurement. However, looking at the changes in the flow pattern in Fig.~\ref{fig:a_dependence_screenshot}, it is clear that the spin must have an impact on the dynamical friction force at some level.

Intuitively, we can understand the total spin-curvature force as arising from two competing effects. Firstly, as the BH travels through the matter, an overdensity of the scalar field forms in the wake of the BH~\cite{Traykova:2021dua}. This overdensity results in the dynamical friction force $F_x$.
In the case of a spinning BH, this overdensity is frame-dragged in the direction of the spin, resulting in a build-up of the field on the counterrotating (negative $y$) side of the flow. This can be clearly seen in the density profiles in Fig.~\ref{fig:a_dependence_screenshot}. Overdense regions tend to gravitationally attract the BH, or, more properly, they contribute to the Weyl curvature that tends to divert the course of the BH in that direction. This first contribution can therefore be identified with the Weyl component of the spin-curvature force, that acts in the negative $y$ direction.
We note that as the velocity increases, overdense features in front of the BH shift in the positive $y$ direction, which may contribute to the fact that the source part of the force appears to be saturating and possibly even reversing at higher $v$, as shown in Fig.~\ref{fig:trends}.

The second contribution to the force comes from $\mathcal{F}$, the momentum flux of the scalar field onto the BH. This tends to act in the opposite direction: the scalar is being accreted preferentially on the counterrotating side, and the flux carries momentum in the positive $y$ direction into the BH. This second contribution can be identified with the Magnus component of the spin-curvature force, since it acts locally on the BH. However, for numerical reasons, we measure this at $r_{\rm i} = 5M$, rather than at the BH horizon, and as a result what we measure contains contributions from the Weyl part too. We clearly see that taking the flux at smaller $r$ makes $\mathcal{F}$ more negative, and expect that it should be possible to reconcile this better with the analytic Magnus force in coordinates such as Kerr-Schild where we are able to measure the flux on the horizon. However, the values shown for $\mathcal{F}$ in Fig.~\ref{fig:force components} give an approximate measure of how this component scales with $v$.

Finally, we note that another limitation of our study is the large uncertainty in the extracted force. 
We use the time evolution of the scalar field to arrive at a steady-state field configurations around the central BH from (arbitrarily chosen) homogeneous initial conditions.  The transient oscillations of the scalar field are long-lived, and so it would be desirable to start with a configuration closer to the final state.

\section{Discussion}
\label{sec:discussion}

In this work we simulated a Kerr BH moving at relativistic velocities through scalar DM that is at rest. We studied the effect of two parameters (the BH spin parameter $a$ and the BH velocity $v$) on the total force transverse to the motion of the BH and to its spin, which we identify with the total spin-curvature force.

We confirmed that the total force on the BH scales linearly with the spin parameter $a$ of the BH up to $a/M = 0.99$, and measured its dependence on the speed $v$ of the BH in the range $0.1 \le v \le 0.55$ for a fixed spin. The behavior at larger $v$ is nonlinear, and so higher-order corrections in the velocity must be important, motivating a search for higher-order analytic results than the previous 1PN estimates that predicted zero net force.

We find that in all cases the total net force is in the opposite direction to the hydrodynamical analogue, although we confirm that it is made up of both positive and negative contributions, and at low speeds it appears to approach the expectation that these Weyl and Magnus components cancel. 
Spin-curvature effects may leave an imprint on gravitational wave signals from extreme mass-ratio inspirals, where the secondary BH has a nonnegligible spin and moves in the presence of a DM cloud. We hope that our simulations can be used to support and extend the limits of analytic results, which are necessary to better quantify such effects in relativistic cases. 
Further studies can also investigate the change in the Magnus force induced by scalar fields with different masses. 

The extracted Magnus force in this paper has large uncertainties due to the long-lived transient oscillations, and we were limited in the duration of the simulations and the range of $v$ that we could explore due to numerical constraints. Using a horizon-penetrating coordinate system that admits a concise analytic expression for the spinning boosted case, such as Cartesian Kerr-Schild coordinates, would help significantly in improving the efficiency and accuracy of the simulations. This is not as straightforward as simply changing the fixed background, as one needs to work out the correct formalism for calculating the exchange of momentum according to the correct asymptotic observers. However, there is no major conceptual or technical barrier to doing so, and we plan to implement this in future work.

While this work was in preparation, we became aware of a concurrent effort to study the aerodynamics of spinning BHs in full general relativity with a semianalytical approach~\cite{Dyson:2024qrq}. This work considers both particle-like and scalar field DM models, and calculates the backreaction force in both the $y$ direction (the direction of the BH spin) and in the $z$ direction (perpendicular to both the BH spin and the BH velocity), finding that the Magnus force reverses direction when the BH velocity is larger than $0.55$. We find tentative agreement with their result in our parameter regime, but we leave a full comparative analysis for future work.

\acknowledgments 

We thank Rodrigo Vicente, Pedro Ferreira and Lam Hui for helpful conversations and input to our preceding works on dynamical friction, which provided the basis for this study. We thank Conor Dyson and Jaime Redondo Yuste for helpful conversations on the effect. We thank the GRTL collaboration (\href{https://github.com/GRTLCollaboration}{https://github.com/GRTLCollaboration}) for their support and code development work, in particular those who worked on the \texttt{GRDzhadzha} code.
E.B. and Z.W. are supported by NSF Grants No. AST-2006538, PHY-2207502, PHY-090003 and PHY-20043, by NASA Grants No. 20-LPS20-0011 and 21-ATP21-0010, by the John Templeton Foundation Grant 62840, by the Simons Foundation, and by the Italian Ministry of Foreign Affairs and International Cooperation Grant No.~PGR01167.
T.H. was supported by NSF grant NSF-1759835.
K.C. acknowledges funding from the UKRI Ernest Rutherford Fellowship (grant number ST/V003240/1).
The simulation work was carried out at the Advanced Research Computing at Hopkins (ARCH) core facility (\url{rockfish.jhu.edu}), which is supported by the NSF Grant No.~OAC-1920103.
The authors also acknowledge the Texas Advanced Computing Center (TACC) at The University of Texas at Austin and the the Sakura
cluster at the Max Planck Computing and Data Facility
(MPCDF) in Garching for providing {HPC, visualization, database, or grid} resources that have contributed to the research results reported within this paper~\cite{10.1145/3311790.3396656}. URL: \url{http://www.tacc.utexas.edu}.

\appendix
\section{Numerical validations}
\label{app:validation}

\begin{figure}[h]
  \includegraphics[width=0.48\textwidth]{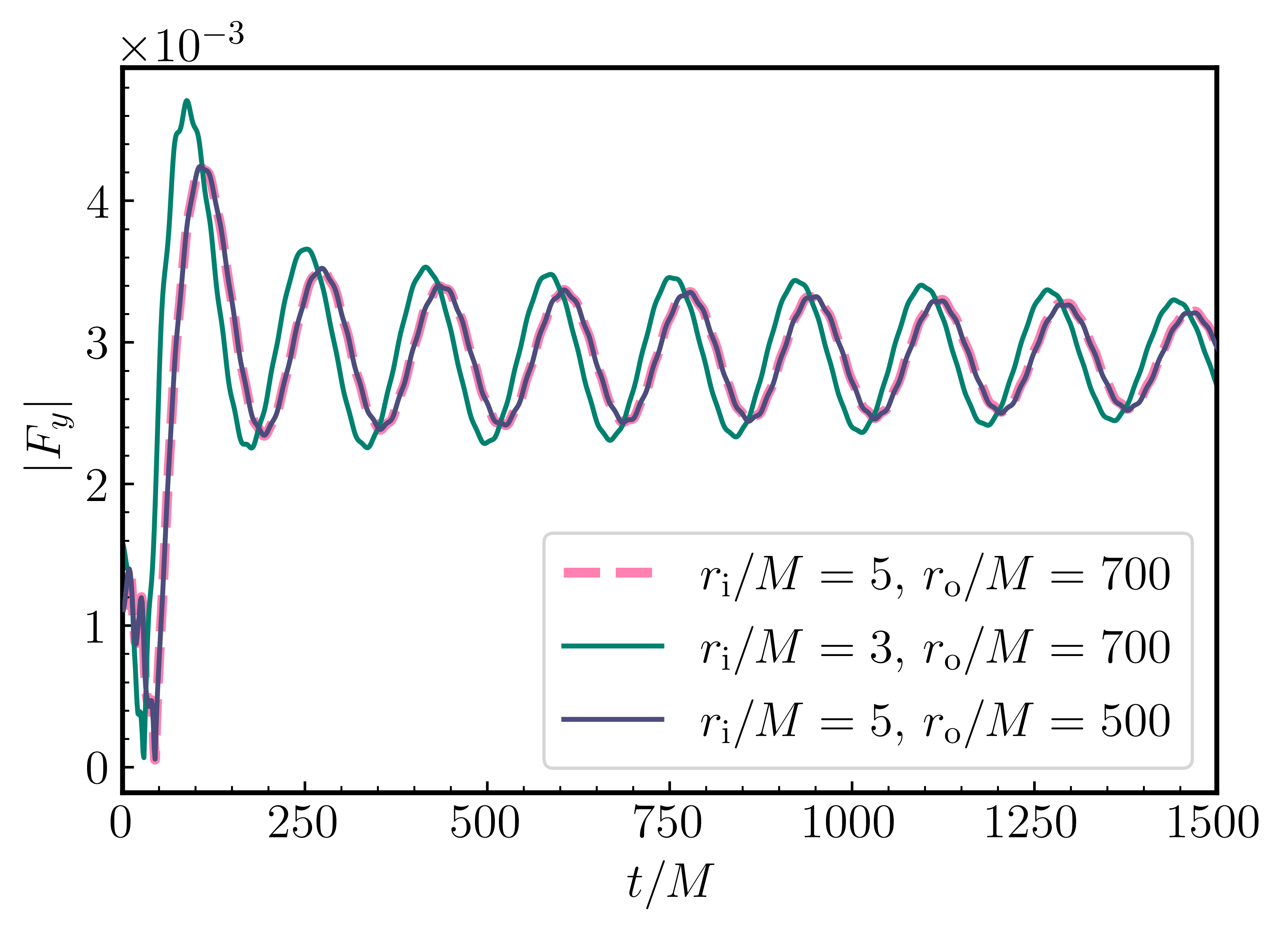}
  \caption{
  The Magnus force on the BH, $F_y$, extracted with different values of $r_{\rm i}$ and $r_{\rm o}$, as a function of simulation time $t/M$. We show here the simulations with parameters $v=0.55$ and $a = 0.7$. The $r_{\rm i}/M=5$, $r_{\rm o} = 700$ line is used to extract the Magnus force magnitude presented in the main text. 
  }
  \label{fig:radius_check}
\end{figure}


\begin{figure}[t]
  \includegraphics[width=0.48\textwidth]{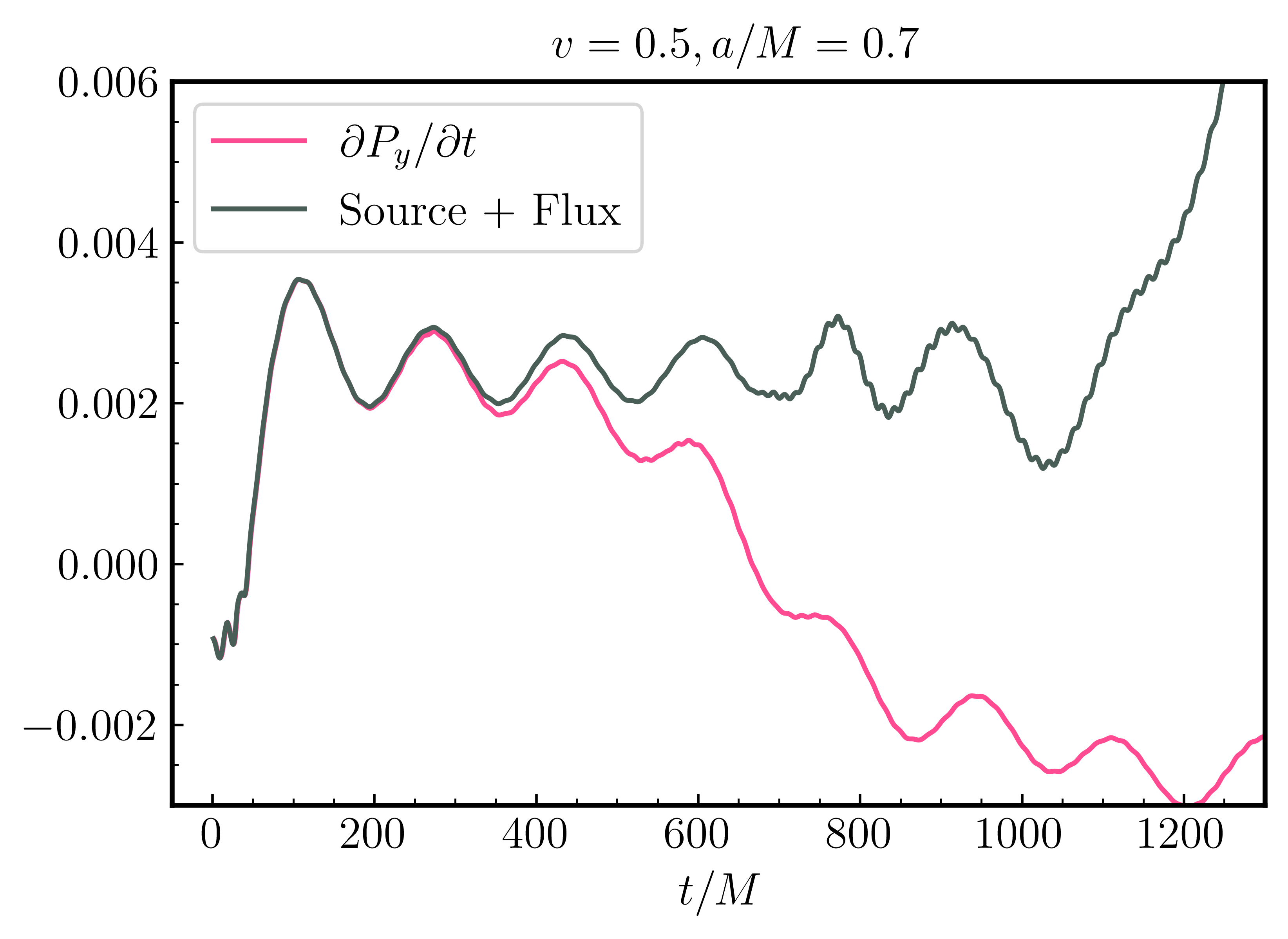}
  \caption{
    This shows our check in which we reconcile the change in the total $y$ momentum within the coordinate volume of interest to the (inner and outer) fluxes at the boundaries, and the source term $\mathcal{S}$, as suggested in Ref.~\cite{Clough:2019jpm}. The total source and flux measurements agree well with the change in momentum at early times, but diverge after around $t=400M$ due to boundary and resolution effects. This is sufficient for us to make a confident measurement of the force, but limits the duration of the simulation in cases of higher $v$ values.
  }
  \label{fig:continuity}
\end{figure}


\begin{figure}[h]
  \includegraphics[width=0.48\textwidth]{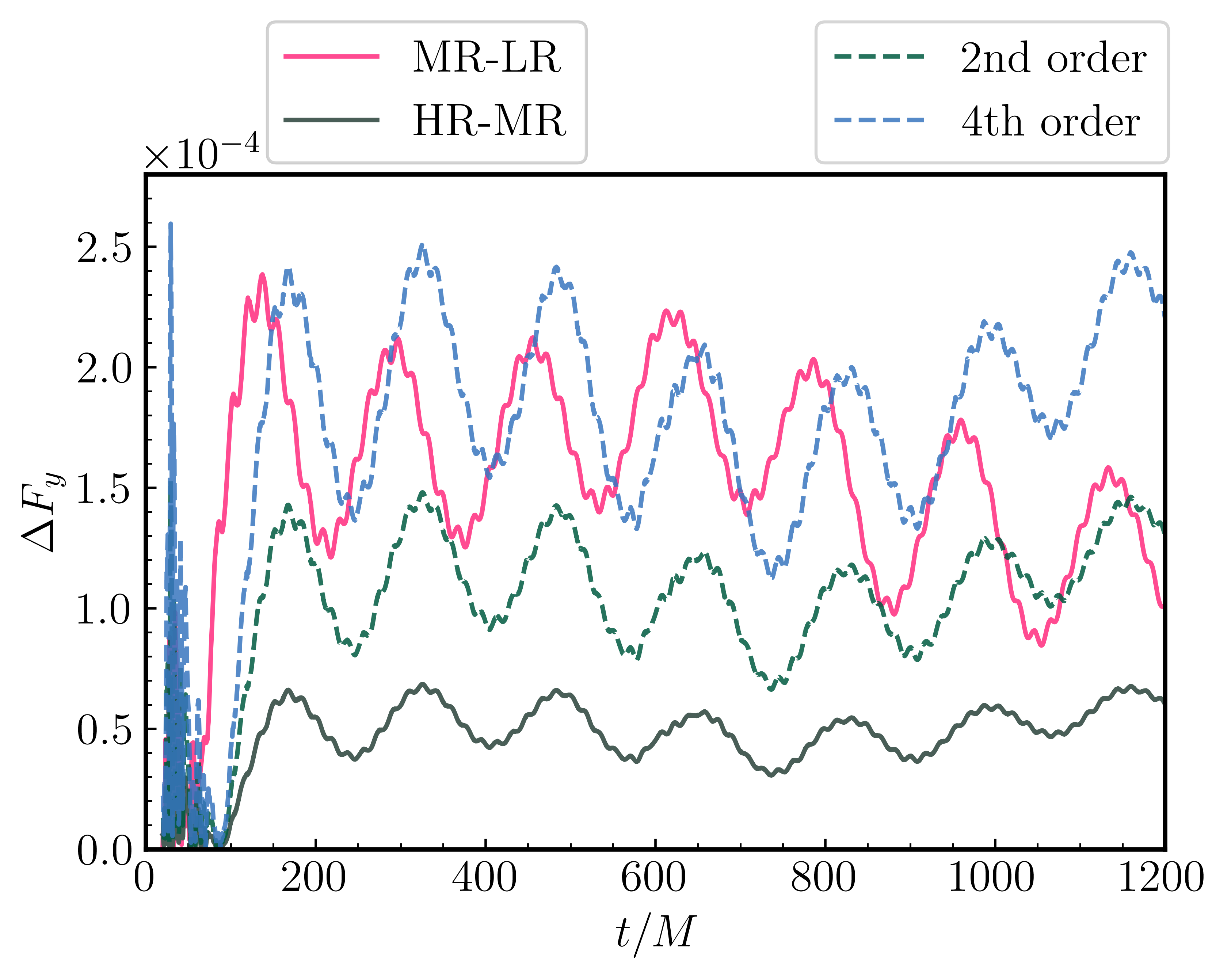}
  \caption{
  The difference in the extracted Magnus force between simulations with three different resolutions, $N_1 = 32$ (LR), $N_2 = 64$ (MR), and $N_3 = 96$ (HR). The black solid line is the difference of the Magnus force between the HR and the MR simulation, and the red solid line is that between the MR and the LR simulation. The dashed green and blue lines are the expected "MR-LR" difference for 2nd and 4th order convergence, respectively. This convergence test is done with parameters $a=0.7$ and $v=0.55$.
  }
  \label{fig:convergence}
  
\end{figure}


We validated our fixed background metric by checking that the Hamiltonian constraints converges to zero at an appropriate order with increasing resolution, and that the time derivatives for the metric fields are zero.

To test the robustness of the spin-curvature force extracted from the simulations to changes in the integration volume, we compared the magnitude of the total force with different radii $r_{\rm i}$, and $r_{\rm o}$ of the extraction surfaces $\Sigma_{\rm i}$, and $\Sigma_{\rm o}$, respectively. The result is shown in Fig.~\ref{fig:radius_check}. 

The Magnus forces extracted with different radii $r_{\rm i}$, $r_{\rm o}$ are found to be similar. The size of the extraction surfaces does slightly affect the value of the Magnus force as a function of the simulation time, but taking the oscillations of the force magnitude as the level of uncertainty (as we did in Fig.~\ref{fig:trends}), the end values of the Magnus force are indistinguishable. Therefore, in the main text, we use $r_{\rm i} = 5M$ and $r_{\rm 0} = 700 M$ for all simulations presented.

Another check we performed was provided by computing the change in the total $y$ momentum within the coordinate volume of interest, and reconciling it to the fluxes and sources measured, as described in Ref.~\cite{Clough:2019jpm}. This provides a strong check on the validity of all parts of the computational domain. We show the case for $v=0.5$ in Fig.~\ref{fig:continuity}, which illustrates well why we find it hard to push our simulations to higher velocities. The total source and flux measurements agree well with the change in momentum at early times, but diverge after around $t=400M$. At this time we are able to make a measurement of the force from the inner flux and source terms as shown in Fig.~\ref{fig:force components}, and these remain regular for some time afterwards. However, the outer flux and the volume integral for the momentum are clearly not behaving well, as they are affected by the lower resolution in the outer part of the domain, and the outer boundary effects. Such effects will eventually propagate inwards and spoil the inner flux and source integrals, and in general we should not trust our results after these integrals start to diverge. Above $v=0.6$ the agreement is lost even before a steady state is reached, and so we are not confident in presenting the results we find for such cases.

We also checked the convergence of the measured force with the BH velocity $v=0.3$ and the BH spin parameter $a=0.99$. We choose three different resolutions $N_1 = 48$, $N_2 = 64$ (which is used in all simulations in the main text), and $N_3 = 80$. The expected convergence factor $c(t)$ is
\begin{equation}
\label{eqn: convergence factor}
    \lim_{\Delta \rightarrow 0} c(t) = \frac{\Delta_1^n - \Delta_2^n}{\Delta_2^n -\Delta_3^n} = 
    \begin{cases}
     2.16 & \textrm{for}\quad n=2\,,\\
     3.66 & \textrm{for}\quad n=4\,,
    \end{cases}
\end{equation}
using $\Delta_{\text{1,2,3}} = L/{N_{\text{1,2,3}}}$.
From the simulations we obtain the force $F_y$. We then calculate the difference $\Delta F_y$ between the middle-resolution run and the low-resolution run (MR-LR), and that between the high-resolution run and the middle-resolution run (HR-MR) . We multiply the (MR-LR) result by the two $c(t)$ factors in \eqref{eqn: convergence factor}, and plot them against the (HR-MR) $\Delta F_y$ in Fig.~\ref{fig:convergence}. 

We see that the (HR-MR) result lies between the 2nd order and 4th order prediction, as expected. We note that the integration method in time used in our code, RK4, is a 4th order numerical scheme, and we use 4th order finite difference stencils in space. However, calculations of the force requires surface and volume integrals, which reduce the total order of convergence. 


\section{Correspondence of Weyl and Magnus forces with flux and curvature source terms}
\label{app:correspondence}


In the main text we make the correspondence between the momentum ``flux'' into the black hole and the Magnus force, and the curvature ``source'' term and the Weyl force, as identified in the previous work of Ref.~\cite{Costa:2018gva}. One can show that in the non-relativistic limit, the accretion of linear $y$-momentum onto a sphere around the BH
\begin{equation*}
    F_{\text{acc},y} = \int_{\partial \Sigma_i} dS_j T_y^j,
\end{equation*}
reduces to the PN expression for the Magnus force,
\begin{equation*}
    F_{\text{mag}} = 4\pi J_x S = 4 \pi \rho v a M.
\end{equation*}
The remaining balance of the forces (the source and the Weyl force) must then also correspond in this limit. This can also be inferred because the Weyl force represents the gravitational effects of nonlocal matter acting on the perturber. In the Newtonian picture, this is the integrated gravitational force of the matter distribution around the object. The curvature ``source'' term corresponds to this quantity in the Newtonian limit, that is,
\begin{multline}
\int_{\Sigma} ~ \dd^3x ~\sqrt{-g}\, T^\mu_\nu \,^{(4)}\Gamma^\nu_{\mu y} \approx \\
\int_{\Sigma} ~ \dd^3x ~\alpha T^0_0 \,^{(4)}\Gamma^0_{0 y} \approx \int_{\Sigma} \dd^3x ~ \frac{\rho M y}{r^3},
\end{multline}
where we have used that $\,^{(4)}\Gamma^0_{0 y} \approx \partial_y \alpha /\alpha$ and $\alpha^2 = 1 - 2M/r$.

To show the correspondence of the flux and the Magnus force, we start from the expression for the deflection angle of a null geodesic around a Kerr BH \cite{Boyer:1966qh}
\begin{equation*}
   \delta = \frac{4M}{b} + \frac{4Ma}{b^2},
\end{equation*}
where $M$ is the mass of the BH, $a$ its spin parameter, and $b$ the impact parameter. This formula is valid on the equatorial plane of the BH, in the limit of small $M/b$ and small $a/M$. It is known that deflection reduces by half for nonrelativistic trajectories. We can also extrapolate this formula out of the equatorial plane by projecting the spin into its component perpendicular to the plane described by the incoming trajectory of the particle and the centre of the perturber, giving
\begin{equation*}
   \delta = \frac{2M}{b} + \frac{2Ma\sin{\theta}}{b^2},
\end{equation*}
where $\theta$ is the angle of the incoming particle with respect to the $z$ axis in the $y$-$z$ plane, as shown in Fig.~\ref{fig:cartoon}. 

Solving for $b$ while taking the limit $a/M \ll 1$ we get
\begin{equation*}
   b = \frac{2M}{\delta} \pm a \sin\theta,
\end{equation*}
where the plus sign is taken when the scattering particle is corotating with the BH ($y>0$), and the minus sign is taken when it is counterrotating ($y<0$).

Assuming that all particles with $\delta > \delta_0$ are accreted onto the BH, then all particles within a semicircle with radius $2M/\delta_0 -a$ in the $+y$ direction and all particles within the radius $2M/\delta_0 + a$ in the $-y$ direction are accreted. 
The deflection of the incoming $x$ momentum into the $y$ direction can be estimated as
\begin{equation*}
   J_y = J_x \delta_0 \frac{y}{\sqrt{y^2+z^2}}
\end{equation*}
on the edge of the semicircle of accretion.
In the limit of $a\rightarrow 0$ the contributions of the two sides would cancel out (i.e., there is no overall accretion of $y$ momentum due to the symmetry of the problem). However, with non-zero BH spin, there is an additional annulus of matter being accreted on the counterrotating side with thickness $a$, and an equal reduction of the accretion on the corotating side.
The additional accreted $y$-momentum per unit time is then
\begin{align*}
   F_{\text{acc},y} &= 2 \int_{\text{annulus}}J_y \,dA 
   \\
   &= 2 J_x \delta_0 \int_{\text{annulus}} \frac{r \sin{\theta}}r r\,dr\,d\theta 
   \\
   &= 4 J_x \delta_0 \int_0^\pi \sin \theta \, d\theta \int_{2M/\delta_0}^{2M/\delta_0 + a \sin\theta} r\,dr  
   \\
   &= 4 \pi J_x a M,
\end{align*}
which is the Magnus force in the non relativistic limit if we make the identification $J_x = \rho v$.

 

\newpage 
\bibliography{mybib}

\begin{thebibliography}{64}%
\makeatletter
\providecommand \@ifxundefined [1]{%
 \@ifx{#1\undefined}
}%
\providecommand \@ifnum [1]{%
 \ifnum #1\expandafter \@firstoftwo
 \else \expandafter \@secondoftwo
 \fi
}%
\providecommand \@ifx [1]{%
 \ifx #1\expandafter \@firstoftwo
 \else \expandafter \@secondoftwo
 \fi
}%
\providecommand \natexlab [1]{#1}%
\providecommand \enquote  [1]{``#1''}%
\providecommand \bibnamefont  [1]{#1}%
\providecommand \bibfnamefont [1]{#1}%
\providecommand \citenamefont [1]{#1}%
\providecommand \href@noop [0]{\@secondoftwo}%
\providecommand \href [0]{\begingroup \@sanitize@url \@href}%
\providecommand \@href[1]{\@@startlink{#1}\@@href}%
\providecommand \@@href[1]{\endgroup#1\@@endlink}%
\providecommand \@sanitize@url [0]{\catcode `\\12\catcode `\$12\catcode
  `\&12\catcode `\#12\catcode `\^12\catcode `\_12\catcode `\%12\relax}%
\providecommand \@@startlink[1]{}%
\providecommand \@@endlink[0]{}%
\providecommand \url  [0]{\begingroup\@sanitize@url \@url }%
\providecommand \@url [1]{\endgroup\@href {#1}{\urlprefix }}%
\providecommand \urlprefix  [0]{URL }%
\providecommand \Eprint [0]{\href }%
\providecommand \doibase [0]{https://doi.org/}%
\providecommand \selectlanguage [0]{\@gobble}%
\providecommand \bibinfo  [0]{\@secondoftwo}%
\providecommand \bibfield  [0]{\@secondoftwo}%
\providecommand \translation [1]{[#1]}%
\providecommand \BibitemOpen [0]{}%
\providecommand \bibitemStop [0]{}%
\providecommand \bibitemNoStop [0]{.\EOS\space}%
\providecommand \EOS [0]{\spacefactor3000\relax}%
\providecommand \BibitemShut  [1]{\csname bibitem#1\endcsname}%
\let\auto@bib@innerbib\@empty
\bibitem [{\citenamefont {Tsuji}\ \emph {et~al.}(1985)\citenamefont {Tsuji},
  \citenamefont {Morikawa},\ and\ \citenamefont {Mizuno}}]{Morikawa}%
  \BibitemOpen
  \bibfield  {author} {\bibinfo {author} {\bibfnamefont {Y.}~\bibnamefont
  {Tsuji}}, \bibinfo {author} {\bibfnamefont {Y.}~\bibnamefont {Morikawa}},\
  and\ \bibinfo {author} {\bibfnamefont {O.}~\bibnamefont {Mizuno}},\
  }\bibfield  {title} {\bibinfo {title} {{Experimental Measurement of the
  Magnus Force on a Rotating Sphere at Low Reynolds Numbers}},\ }\href
  {https://doi.org/10.1115/1.3242517} {\bibfield  {journal} {\bibinfo
  {journal} {Journal of Fluids Engineering}\ }\textbf {\bibinfo {volume}
  {107}},\ \bibinfo {pages} {484} (\bibinfo {year} {1985})}\BibitemShut
  {NoStop}%
\bibitem [{\citenamefont {Rubinow}\ and\ \citenamefont
  {Keller}(1961)}]{rubinow_keller_1961}%
  \BibitemOpen
  \bibfield  {author} {\bibinfo {author} {\bibfnamefont {S.~I.}\ \bibnamefont
  {Rubinow}}\ and\ \bibinfo {author} {\bibfnamefont {J.~B.}\ \bibnamefont
  {Keller}},\ }\bibfield  {title} {\bibinfo {title} {The transverse force on a
  spinning sphere moving in a viscous fluid},\ }\href
  {https://doi.org/10.1017/S0022112061000640} {\bibfield  {journal} {\bibinfo
  {journal} {Journal of Fluid Mechanics}\ }\textbf {\bibinfo {volume} {11}},\
  \bibinfo {pages} {447–459} (\bibinfo {year} {1961})}\BibitemShut {NoStop}%
\bibitem [{\citenamefont {Munson}\ \emph {et~al.}(2016)\citenamefont {Munson},
  \citenamefont {Gerhart}, \citenamefont {Gerhart}, \citenamefont {Hochstein},
  \citenamefont {Young},\ and\ \citenamefont {Okiishi}}]{munson_munson_2016}%
  \BibitemOpen
  \bibfield  {author} {\bibinfo {author} {\bibfnamefont {B.~R.}\ \bibnamefont
  {Munson}}, \bibinfo {author} {\bibfnamefont {P.~M.}\ \bibnamefont {Gerhart}},
  \bibinfo {author} {\bibfnamefont {A.~L.}\ \bibnamefont {Gerhart}}, \bibinfo
  {author} {\bibfnamefont {J.~I.}\ \bibnamefont {Hochstein}}, \bibinfo {author}
  {\bibfnamefont {D.~F.}\ \bibnamefont {Young}},\ and\ \bibinfo {author}
  {\bibfnamefont {T.~H.}\ \bibnamefont {Okiishi}},\ }\href@noop {} {\emph
  {\bibinfo {title} {Munson, {Young}, and {Okiishi}'s {Fundamentals} of fluid
  mechanics}}},\ \bibinfo {edition} {eighth edition}\ ed.\ (\bibinfo
  {publisher} {Wiley Hoboken, NJ},\ \bibinfo {address} {Hoboken, NJ},\ \bibinfo
  {year} {2016})\ \bibinfo {note} {section: xxii, 777, 11 pages : illustrations
  ; 28 cm}\BibitemShut {NoStop}%
\bibitem [{\citenamefont {Font}\ \emph {et~al.}(1999)\citenamefont {Font},
  \citenamefont {Ibanez},\ and\ \citenamefont {Papadopoulos}}]{Font:1998sc}%
  \BibitemOpen
  \bibfield  {author} {\bibinfo {author} {\bibfnamefont {J.~A.}\ \bibnamefont
  {Font}}, \bibinfo {author} {\bibfnamefont {J.~M.}\ \bibnamefont {Ibanez}},\
  and\ \bibinfo {author} {\bibfnamefont {P.}~\bibnamefont {Papadopoulos}},\
  }\bibfield  {title} {\bibinfo {title} {{Nonaxisymmetric relativistic
  Bondi-Hoyle accretion onto a Kerr black hole}},\ }\href
  {https://doi.org/10.1046/j.1365-8711.1999.02459.x} {\bibfield  {journal}
  {\bibinfo  {journal} {Mon. Not. Roy. Astron. Soc.}\ }\textbf {\bibinfo
  {volume} {305}},\ \bibinfo {pages} {920} (\bibinfo {year} {1999})},\ \Eprint
  {https://arxiv.org/abs/astro-ph/9810344} {arXiv:astro-ph/9810344}
  \BibitemShut {NoStop}%
\bibitem [{\citenamefont {Okawa}\ and\ \citenamefont
  {Cardoso}(2014)}]{Okawa:2014sxa}%
  \BibitemOpen
  \bibfield  {author} {\bibinfo {author} {\bibfnamefont {H.}~\bibnamefont
  {Okawa}}\ and\ \bibinfo {author} {\bibfnamefont {V.}~\bibnamefont
  {Cardoso}},\ }\bibfield  {title} {\bibinfo {title} {{Black holes and
  fundamental fields: Hair, kicks, and a gravitational Magnus effect}},\ }\href
  {https://doi.org/10.1103/PhysRevD.90.104040} {\bibfield  {journal} {\bibinfo
  {journal} {Phys. Rev. D}\ }\textbf {\bibinfo {volume} {90}},\ \bibinfo
  {pages} {104040} (\bibinfo {year} {2014})},\ \Eprint
  {https://arxiv.org/abs/1405.4861} {arXiv:1405.4861 [gr-qc]} \BibitemShut
  {NoStop}%
\bibitem [{\citenamefont {Cashen}\ \emph {et~al.}(2017)\citenamefont {Cashen},
  \citenamefont {Aker},\ and\ \citenamefont {Kesden}}]{Cashen:2016neh}%
  \BibitemOpen
  \bibfield  {author} {\bibinfo {author} {\bibfnamefont {B.}~\bibnamefont
  {Cashen}}, \bibinfo {author} {\bibfnamefont {A.}~\bibnamefont {Aker}},\ and\
  \bibinfo {author} {\bibfnamefont {M.}~\bibnamefont {Kesden}},\ }\bibfield
  {title} {\bibinfo {title} {{Gravitomagnetic dynamical friction}},\ }\href
  {https://doi.org/10.1103/PhysRevD.95.064014} {\bibfield  {journal} {\bibinfo
  {journal} {Phys. Rev. D}\ }\textbf {\bibinfo {volume} {95}},\ \bibinfo
  {pages} {064014} (\bibinfo {year} {2017})},\ \Eprint
  {https://arxiv.org/abs/1610.01590} {arXiv:1610.01590 [gr-qc]} \BibitemShut
  {NoStop}%
\bibitem [{\citenamefont {Costa}\ \emph {et~al.}(2018)\citenamefont {Costa},
  \citenamefont {Franco},\ and\ \citenamefont {Cardoso}}]{Costa:2018gva}%
  \BibitemOpen
  \bibfield  {author} {\bibinfo {author} {\bibfnamefont {L.~F.~O.}\
  \bibnamefont {Costa}}, \bibinfo {author} {\bibfnamefont {R.}~\bibnamefont
  {Franco}},\ and\ \bibinfo {author} {\bibfnamefont {V.}~\bibnamefont
  {Cardoso}},\ }\bibfield  {title} {\bibinfo {title} {{Gravitational Magnus
  effect}},\ }\href {https://doi.org/10.1103/PhysRevD.98.024026} {\bibfield
  {journal} {\bibinfo  {journal} {Phys. Rev. D}\ }\textbf {\bibinfo {volume}
  {98}},\ \bibinfo {pages} {024026} (\bibinfo {year} {2018})},\ \Eprint
  {https://arxiv.org/abs/1805.01097} {arXiv:1805.01097 [gr-qc]} \BibitemShut
  {NoStop}%
\bibitem [{\citenamefont {Amaro-Seoane}(2018)}]{Amaro-Seoane:2012lgq}%
  \BibitemOpen
  \bibfield  {author} {\bibinfo {author} {\bibfnamefont {P.}~\bibnamefont
  {Amaro-Seoane}},\ }\bibfield  {title} {\bibinfo {title} {{Relativistic
  dynamics and extreme mass ratio inspirals}},\ }\href
  {https://doi.org/10.1007/s41114-018-0013-8} {\bibfield  {journal} {\bibinfo
  {journal} {Living Rev. Rel.}\ }\textbf {\bibinfo {volume} {21}},\ \bibinfo
  {pages} {4} (\bibinfo {year} {2018})},\ \Eprint
  {https://arxiv.org/abs/1205.5240} {arXiv:1205.5240 [astro-ph.CO]}
  \BibitemShut {NoStop}%
\bibitem [{\citenamefont {Afshordi}\ \emph {et~al.}(2023)\citenamefont
  {Afshordi} \emph {et~al.}}]{LISAConsortiumWaveformWorkingGroup:2023arg}%
  \BibitemOpen
  \bibfield  {author} {\bibinfo {author} {\bibfnamefont {N.}~\bibnamefont
  {Afshordi}} \emph {et~al.} (\bibinfo {collaboration} {LISA Consortium
  Waveform Working Group}),\ }\bibfield  {title} {\bibinfo {title} {{Waveform
  Modelling for the Laser Interferometer Space Antenna}},\ }\href@noop {} {\
  (\bibinfo {year} {2023})},\ \Eprint {https://arxiv.org/abs/2311.01300}
  {arXiv:2311.01300 [gr-qc]} \BibitemShut {NoStop}%
\bibitem [{\citenamefont {Berry}\ \emph {et~al.}(2019)\citenamefont {Berry},
  \citenamefont {Hughes}, \citenamefont {Sopuerta}, \citenamefont {Chua},
  \citenamefont {Heffernan}, \citenamefont {Holley-Bockelmann}, \citenamefont
  {Mihaylov}, \citenamefont {Miller},\ and\ \citenamefont
  {Sesana}}]{Berry:2019wgg}%
  \BibitemOpen
  \bibfield  {author} {\bibinfo {author} {\bibfnamefont {C.~P.~L.}\
  \bibnamefont {Berry}}, \bibinfo {author} {\bibfnamefont {S.~A.}\ \bibnamefont
  {Hughes}}, \bibinfo {author} {\bibfnamefont {C.~F.}\ \bibnamefont
  {Sopuerta}}, \bibinfo {author} {\bibfnamefont {A.~J.~K.}\ \bibnamefont
  {Chua}}, \bibinfo {author} {\bibfnamefont {A.}~\bibnamefont {Heffernan}},
  \bibinfo {author} {\bibfnamefont {K.}~\bibnamefont {Holley-Bockelmann}},
  \bibinfo {author} {\bibfnamefont {D.~P.}\ \bibnamefont {Mihaylov}}, \bibinfo
  {author} {\bibfnamefont {M.~C.}\ \bibnamefont {Miller}},\ and\ \bibinfo
  {author} {\bibfnamefont {A.}~\bibnamefont {Sesana}},\ }\bibfield  {title}
  {\bibinfo {title} {{The unique potential of extreme mass-ratio inspirals for
  gravitational-wave astronomy}},\ }\href@noop {} {\  (\bibinfo {year}
  {2019})},\ \Eprint {https://arxiv.org/abs/1903.03686} {arXiv:1903.03686
  [astro-ph.HE]} \BibitemShut {NoStop}%
\bibitem [{\citenamefont {Hannuksela}\ \emph {et~al.}(2019)\citenamefont
  {Hannuksela}, \citenamefont {Wong}, \citenamefont {Brito}, \citenamefont
  {Berti},\ and\ \citenamefont {Li}}]{Hannuksela:2018izj}%
  \BibitemOpen
  \bibfield  {author} {\bibinfo {author} {\bibfnamefont {O.~A.}\ \bibnamefont
  {Hannuksela}}, \bibinfo {author} {\bibfnamefont {K.~W.~K.}\ \bibnamefont
  {Wong}}, \bibinfo {author} {\bibfnamefont {R.}~\bibnamefont {Brito}},
  \bibinfo {author} {\bibfnamefont {E.}~\bibnamefont {Berti}},\ and\ \bibinfo
  {author} {\bibfnamefont {T.~G.~F.}\ \bibnamefont {Li}},\ }\bibfield  {title}
  {\bibinfo {title} {{Probing the existence of ultralight bosons with a single
  gravitational-wave measurement}},\ }\href
  {https://doi.org/10.1038/s41550-019-0712-4} {\bibfield  {journal} {\bibinfo
  {journal} {Nature Astron.}\ }\textbf {\bibinfo {volume} {3}},\ \bibinfo
  {pages} {447} (\bibinfo {year} {2019})},\ \Eprint
  {https://arxiv.org/abs/1804.09659} {arXiv:1804.09659 [astro-ph.HE]}
  \BibitemShut {NoStop}%
\bibitem [{\citenamefont {Traykova}\ \emph {et~al.}(2021)\citenamefont
  {Traykova}, \citenamefont {Clough}, \citenamefont {Helfer}, \citenamefont
  {Berti}, \citenamefont {Ferreira},\ and\ \citenamefont
  {Hui}}]{Traykova:2021dua}%
  \BibitemOpen
  \bibfield  {author} {\bibinfo {author} {\bibfnamefont {D.}~\bibnamefont
  {Traykova}}, \bibinfo {author} {\bibfnamefont {K.}~\bibnamefont {Clough}},
  \bibinfo {author} {\bibfnamefont {T.}~\bibnamefont {Helfer}}, \bibinfo
  {author} {\bibfnamefont {E.}~\bibnamefont {Berti}}, \bibinfo {author}
  {\bibfnamefont {P.~G.}\ \bibnamefont {Ferreira}},\ and\ \bibinfo {author}
  {\bibfnamefont {L.}~\bibnamefont {Hui}},\ }\bibfield  {title} {\bibinfo
  {title} {{Dynamical friction from scalar dark matter in the relativistic
  regime}},\ }\href {https://doi.org/10.1103/PhysRevD.104.103014} {\bibfield
  {journal} {\bibinfo  {journal} {Phys. Rev. D}\ }\textbf {\bibinfo {volume}
  {104}},\ \bibinfo {pages} {103014} (\bibinfo {year} {2021})},\ \Eprint
  {https://arxiv.org/abs/2106.08280} {arXiv:2106.08280 [gr-qc]} \BibitemShut
  {NoStop}%
\bibitem [{\citenamefont {Traykova}\ \emph {et~al.}(2023)\citenamefont
  {Traykova}, \citenamefont {Vicente}, \citenamefont {Clough}, \citenamefont
  {Helfer}, \citenamefont {Berti}, \citenamefont {Ferreira},\ and\
  \citenamefont {Hui}}]{Traykova:2023qyv}%
  \BibitemOpen
  \bibfield  {author} {\bibinfo {author} {\bibfnamefont {D.}~\bibnamefont
  {Traykova}}, \bibinfo {author} {\bibfnamefont {R.}~\bibnamefont {Vicente}},
  \bibinfo {author} {\bibfnamefont {K.}~\bibnamefont {Clough}}, \bibinfo
  {author} {\bibfnamefont {T.}~\bibnamefont {Helfer}}, \bibinfo {author}
  {\bibfnamefont {E.}~\bibnamefont {Berti}}, \bibinfo {author} {\bibfnamefont
  {P.~G.}\ \bibnamefont {Ferreira}},\ and\ \bibinfo {author} {\bibfnamefont
  {L.}~\bibnamefont {Hui}},\ }\bibfield  {title} {\bibinfo {title}
  {{Relativistic drag forces on black holes from scalar dark matter clouds of
  all sizes}},\ }\href {https://doi.org/10.1103/PhysRevD.108.L121502}
  {\bibfield  {journal} {\bibinfo  {journal} {Phys. Rev. D}\ }\textbf {\bibinfo
  {volume} {108}},\ \bibinfo {pages} {L121502} (\bibinfo {year} {2023})},\
  \Eprint {https://arxiv.org/abs/2305.10492} {arXiv:2305.10492 [gr-qc]}
  \BibitemShut {NoStop}%
\bibitem [{\citenamefont {Vicente}\ and\ \citenamefont
  {Cardoso}(2022)}]{Vicente:2022ivh}%
  \BibitemOpen
  \bibfield  {author} {\bibinfo {author} {\bibfnamefont {R.}~\bibnamefont
  {Vicente}}\ and\ \bibinfo {author} {\bibfnamefont {V.}~\bibnamefont
  {Cardoso}},\ }\bibfield  {title} {\bibinfo {title} {{Dynamical friction of
  black holes in ultralight dark matter}},\ }\href
  {https://doi.org/10.1103/PhysRevD.105.083008} {\bibfield  {journal} {\bibinfo
   {journal} {Phys. Rev. D}\ }\textbf {\bibinfo {volume} {105}},\ \bibinfo
  {pages} {083008} (\bibinfo {year} {2022})},\ \Eprint
  {https://arxiv.org/abs/2201.08854} {arXiv:2201.08854 [gr-qc]} \BibitemShut
  {NoStop}%
\bibitem [{\citenamefont {Barausse}\ \emph {et~al.}(2014)\citenamefont
  {Barausse}, \citenamefont {Cardoso},\ and\ \citenamefont
  {Pani}}]{Barausse:2014tra}%
  \BibitemOpen
  \bibfield  {author} {\bibinfo {author} {\bibfnamefont {E.}~\bibnamefont
  {Barausse}}, \bibinfo {author} {\bibfnamefont {V.}~\bibnamefont {Cardoso}},\
  and\ \bibinfo {author} {\bibfnamefont {P.}~\bibnamefont {Pani}},\ }\bibfield
  {title} {\bibinfo {title} {{Can environmental effects spoil precision
  gravitational-wave astrophysics?}},\ }\href
  {https://doi.org/10.1103/PhysRevD.89.104059} {\bibfield  {journal} {\bibinfo
  {journal} {Phys. Rev. D}\ }\textbf {\bibinfo {volume} {89}},\ \bibinfo
  {pages} {104059} (\bibinfo {year} {2014})},\ \Eprint
  {https://arxiv.org/abs/1404.7149} {arXiv:1404.7149 [gr-qc]} \BibitemShut
  {NoStop}%
\bibitem [{\citenamefont {Bonga}\ \emph {et~al.}(2019)\citenamefont {Bonga},
  \citenamefont {Yang},\ and\ \citenamefont {Hughes}}]{Bonga:2019ycj}%
  \BibitemOpen
  \bibfield  {author} {\bibinfo {author} {\bibfnamefont {B.}~\bibnamefont
  {Bonga}}, \bibinfo {author} {\bibfnamefont {H.}~\bibnamefont {Yang}},\ and\
  \bibinfo {author} {\bibfnamefont {S.~A.}\ \bibnamefont {Hughes}},\ }\bibfield
   {title} {\bibinfo {title} {{Tidal resonance in extreme mass-ratio
  inspirals}},\ }\href {https://doi.org/10.1103/PhysRevLett.123.101103}
  {\bibfield  {journal} {\bibinfo  {journal} {Phys. Rev. Lett.}\ }\textbf
  {\bibinfo {volume} {123}},\ \bibinfo {pages} {101103} (\bibinfo {year}
  {2019})},\ \Eprint {https://arxiv.org/abs/1905.00030} {arXiv:1905.00030
  [gr-qc]} \BibitemShut {NoStop}%
\bibitem [{\citenamefont {Nesti}\ and\ \citenamefont
  {Salucci}(2013)}]{Nesti:2013uwa}%
  \BibitemOpen
  \bibfield  {author} {\bibinfo {author} {\bibfnamefont {F.}~\bibnamefont
  {Nesti}}\ and\ \bibinfo {author} {\bibfnamefont {P.}~\bibnamefont
  {Salucci}},\ }\bibfield  {title} {\bibinfo {title} {{The Dark Matter halo of
  the Milky Way, AD 2013}},\ }\href
  {https://doi.org/10.1088/1475-7516/2013/07/016} {\bibfield  {journal}
  {\bibinfo  {journal} {JCAP}\ }\textbf {\bibinfo {volume} {07}},\ \bibinfo
  {pages} {016}},\ \Eprint {https://arxiv.org/abs/1304.5127} {arXiv:1304.5127
  [astro-ph.GA]} \BibitemShut {NoStop}%
\bibitem [{\citenamefont {Pato}\ \emph {et~al.}(2015)\citenamefont {Pato},
  \citenamefont {Iocco},\ and\ \citenamefont {Bertone}}]{Pato:2015dua}%
  \BibitemOpen
  \bibfield  {author} {\bibinfo {author} {\bibfnamefont {M.}~\bibnamefont
  {Pato}}, \bibinfo {author} {\bibfnamefont {F.}~\bibnamefont {Iocco}},\ and\
  \bibinfo {author} {\bibfnamefont {G.}~\bibnamefont {Bertone}},\ }\bibfield
  {title} {\bibinfo {title} {{Dynamical constraints on the dark matter
  distribution in the Milky Way}},\ }\href
  {https://doi.org/10.1088/1475-7516/2015/12/001} {\bibfield  {journal}
  {\bibinfo  {journal} {JCAP}\ }\textbf {\bibinfo {volume} {12}},\ \bibinfo
  {pages} {001}},\ \Eprint {https://arxiv.org/abs/1504.06324} {arXiv:1504.06324
  [astro-ph.GA]} \BibitemShut {NoStop}%
\bibitem [{\citenamefont {De~Martino}\ \emph {et~al.}(2020)\citenamefont
  {De~Martino}, \citenamefont {Broadhurst}, \citenamefont {Tye}, \citenamefont
  {Chiueh},\ and\ \citenamefont {Schive}}]{DeMartino:2018zkx}%
  \BibitemOpen
  \bibfield  {author} {\bibinfo {author} {\bibfnamefont {I.}~\bibnamefont
  {De~Martino}}, \bibinfo {author} {\bibfnamefont {T.}~\bibnamefont
  {Broadhurst}}, \bibinfo {author} {\bibfnamefont {S.~H.~H.}\ \bibnamefont
  {Tye}}, \bibinfo {author} {\bibfnamefont {T.}~\bibnamefont {Chiueh}},\ and\
  \bibinfo {author} {\bibfnamefont {H.-Y.}\ \bibnamefont {Schive}},\ }\bibfield
   {title} {\bibinfo {title} {{Dynamical Evidence of a Solitonic Core of
  $10^{9}M_\odot$ in the Milky Way}},\ }\href
  {https://doi.org/10.1016/j.dark.2020.100503} {\bibfield  {journal} {\bibinfo
  {journal} {Phys. Dark Univ.}\ }\textbf {\bibinfo {volume} {28}},\ \bibinfo
  {pages} {100503} (\bibinfo {year} {2020})},\ \Eprint
  {https://arxiv.org/abs/1807.08153} {arXiv:1807.08153 [astro-ph.GA]}
  \BibitemShut {NoStop}%
\bibitem [{\citenamefont {Li}\ \emph {et~al.}(2020)\citenamefont {Li},
  \citenamefont {Shen},\ and\ \citenamefont {Schive}}]{Li:2020qva}%
  \BibitemOpen
  \bibfield  {author} {\bibinfo {author} {\bibfnamefont {Z.}~\bibnamefont
  {Li}}, \bibinfo {author} {\bibfnamefont {J.}~\bibnamefont {Shen}},\ and\
  \bibinfo {author} {\bibfnamefont {H.-Y.}\ \bibnamefont {Schive}},\ }\bibfield
   {title} {\bibinfo {title} {{Testing the Prediction of Fuzzy Dark Matter
  Theory in the Milky Way Center}}\ }\href
  {https://doi.org/10.3847/1538-4357/ab6598} {10.3847/1538-4357/ab6598}
  (\bibinfo {year} {2020}),\ \Eprint {https://arxiv.org/abs/2001.00318}
  {arXiv:2001.00318 [astro-ph.GA]} \BibitemShut {NoStop}%
\bibitem [{\citenamefont {Ablimit}\ \emph {et~al.}(2020)\citenamefont
  {Ablimit}, \citenamefont {Zhao}, \citenamefont {Flynn},\ and\ \citenamefont
  {Bird}}]{Ablimit:2020gxw}%
  \BibitemOpen
  \bibfield  {author} {\bibinfo {author} {\bibfnamefont {I.}~\bibnamefont
  {Ablimit}}, \bibinfo {author} {\bibfnamefont {G.}~\bibnamefont {Zhao}},
  \bibinfo {author} {\bibfnamefont {C.}~\bibnamefont {Flynn}},\ and\ \bibinfo
  {author} {\bibfnamefont {S.~A.}\ \bibnamefont {Bird}},\ }\bibfield  {title}
  {\bibinfo {title} {{The Rotation Curve, Mass Distribution and Dark Matter
  Content of the Milky Way from Classical Cepheids}},\ }\href
  {https://doi.org/10.3847/2041-8213/ab8d45} {\bibfield  {journal} {\bibinfo
  {journal} {Astrophys. J.}\ }\textbf {\bibinfo {volume} {895}},\ \bibinfo
  {pages} {L12} (\bibinfo {year} {2020})},\ \Eprint
  {https://arxiv.org/abs/2004.13768} {arXiv:2004.13768 [astro-ph.GA]}
  \BibitemShut {NoStop}%
\bibitem [{\citenamefont {{Zel'Dovich}}(1971)}]{1971JETPL..14..180Z}%
  \BibitemOpen
  \bibfield  {author} {\bibinfo {author} {\bibfnamefont {Y.~B.}\ \bibnamefont
  {{Zel'Dovich}}},\ }\bibfield  {title} {\bibinfo {title} {{Generation of Waves
  by a Rotating Body}},\ }\href@noop {} {\bibfield  {journal} {\bibinfo
  {journal} {Soviet Journal of Experimental and Theoretical Physics Letters}\
  }\textbf {\bibinfo {volume} {14}},\ \bibinfo {pages} {180} (\bibinfo {year}
  {1971})}\BibitemShut {NoStop}%
\bibitem [{\citenamefont {Press}\ and\ \citenamefont
  {Teukolsky}(1972)}]{Press:1972zz}%
  \BibitemOpen
  \bibfield  {author} {\bibinfo {author} {\bibfnamefont {W.~H.}\ \bibnamefont
  {Press}}\ and\ \bibinfo {author} {\bibfnamefont {S.~A.}\ \bibnamefont
  {Teukolsky}},\ }\bibfield  {title} {\bibinfo {title} {{Floating Orbits,
  Superradiant Scattering and the Black-hole Bomb}},\ }\href
  {https://doi.org/10.1038/238211a0} {\bibfield  {journal} {\bibinfo  {journal}
  {Nature}\ }\textbf {\bibinfo {volume} {238}},\ \bibinfo {pages} {211}
  (\bibinfo {year} {1972})}\BibitemShut {NoStop}%
\bibitem [{\citenamefont {Zouros}\ and\ \citenamefont
  {Eardley}(1979)}]{Zouros:1979iw}%
  \BibitemOpen
  \bibfield  {author} {\bibinfo {author} {\bibfnamefont {T.~J.~M.}\
  \bibnamefont {Zouros}}\ and\ \bibinfo {author} {\bibfnamefont {D.~M.}\
  \bibnamefont {Eardley}},\ }\bibfield  {title} {\bibinfo {title}
  {{INSTABILITIES OF MASSIVE SCALAR PERTURBATIONS OF A ROTATING BLACK HOLE}},\
  }\href {https://doi.org/10.1016/0003-4916(79)90237-9} {\bibfield  {journal}
  {\bibinfo  {journal} {Annals Phys.}\ }\textbf {\bibinfo {volume} {118}},\
  \bibinfo {pages} {139} (\bibinfo {year} {1979})}\BibitemShut {NoStop}%
\bibitem [{\citenamefont {Detweiler}(1980)}]{Detweiler:1980uk}%
  \BibitemOpen
  \bibfield  {author} {\bibinfo {author} {\bibfnamefont {S.~L.}\ \bibnamefont
  {Detweiler}},\ }\bibfield  {title} {\bibinfo {title} {{KLEIN-GORDON EQUATION
  AND ROTATING BLACK HOLES}},\ }\href
  {https://doi.org/10.1103/PhysRevD.22.2323} {\bibfield  {journal} {\bibinfo
  {journal} {Phys. Rev. D}\ }\textbf {\bibinfo {volume} {22}},\ \bibinfo
  {pages} {2323} (\bibinfo {year} {1980})}\BibitemShut {NoStop}%
\bibitem [{\citenamefont {Cardoso}\ \emph {et~al.}(2004)\citenamefont
  {Cardoso}, \citenamefont {Dias}, \citenamefont {Lemos},\ and\ \citenamefont
  {Yoshida}}]{Cardoso:2004nk}%
  \BibitemOpen
  \bibfield  {author} {\bibinfo {author} {\bibfnamefont {V.}~\bibnamefont
  {Cardoso}}, \bibinfo {author} {\bibfnamefont {O.~J.~C.}\ \bibnamefont
  {Dias}}, \bibinfo {author} {\bibfnamefont {J.~P.~S.}\ \bibnamefont {Lemos}},\
  and\ \bibinfo {author} {\bibfnamefont {S.}~\bibnamefont {Yoshida}},\
  }\bibfield  {title} {\bibinfo {title} {{The Black hole bomb and superradiant
  instabilities}},\ }\href {https://doi.org/10.1103/PhysRevD.70.049903}
  {\bibfield  {journal} {\bibinfo  {journal} {Phys. Rev. D}\ }\textbf {\bibinfo
  {volume} {70}},\ \bibinfo {pages} {044039} (\bibinfo {year} {2004})},\
  \bibinfo {note} {[Erratum: Phys.Rev.D 70, 049903 (2004)]},\ \Eprint
  {https://arxiv.org/abs/hep-th/0404096} {arXiv:hep-th/0404096} \BibitemShut
  {NoStop}%
\bibitem [{\citenamefont {Cardoso}\ and\ \citenamefont
  {Yoshida}(2005)}]{Cardoso:2005vk}%
  \BibitemOpen
  \bibfield  {author} {\bibinfo {author} {\bibfnamefont {V.}~\bibnamefont
  {Cardoso}}\ and\ \bibinfo {author} {\bibfnamefont {S.}~\bibnamefont
  {Yoshida}},\ }\bibfield  {title} {\bibinfo {title} {{Superradiant
  instabilities of rotating black branes and strings}},\ }\href
  {https://doi.org/10.1088/1126-6708/2005/07/009} {\bibfield  {journal}
  {\bibinfo  {journal} {JHEP}\ }\textbf {\bibinfo {volume} {07}},\ \bibinfo
  {pages} {009}},\ \Eprint {https://arxiv.org/abs/hep-th/0502206}
  {arXiv:hep-th/0502206} \BibitemShut {NoStop}%
\bibitem [{\citenamefont {Dolan}(2007)}]{Dolan:2007mj}%
  \BibitemOpen
  \bibfield  {author} {\bibinfo {author} {\bibfnamefont {S.~R.}\ \bibnamefont
  {Dolan}},\ }\bibfield  {title} {\bibinfo {title} {{Instability of the massive
  Klein-Gordon field on the Kerr spacetime}},\ }\href
  {https://doi.org/10.1103/PhysRevD.76.084001} {\bibfield  {journal} {\bibinfo
  {journal} {Phys. Rev. D}\ }\textbf {\bibinfo {volume} {76}},\ \bibinfo
  {pages} {084001} (\bibinfo {year} {2007})},\ \Eprint
  {https://arxiv.org/abs/0705.2880} {arXiv:0705.2880 [gr-qc]} \BibitemShut
  {NoStop}%
\bibitem [{\citenamefont {Arvanitaki}\ and\ \citenamefont
  {Dubovsky}(2011)}]{Arvanitaki:2010sy}%
  \BibitemOpen
  \bibfield  {author} {\bibinfo {author} {\bibfnamefont {A.}~\bibnamefont
  {Arvanitaki}}\ and\ \bibinfo {author} {\bibfnamefont {S.}~\bibnamefont
  {Dubovsky}},\ }\bibfield  {title} {\bibinfo {title} {{Exploring the String
  Axiverse with Precision Black Hole Physics}},\ }\href
  {https://doi.org/10.1103/PhysRevD.83.044026} {\bibfield  {journal} {\bibinfo
  {journal} {Phys. Rev. D}\ }\textbf {\bibinfo {volume} {83}},\ \bibinfo
  {pages} {044026} (\bibinfo {year} {2011})},\ \Eprint
  {https://arxiv.org/abs/1004.3558} {arXiv:1004.3558 [hep-th]} \BibitemShut
  {NoStop}%
\bibitem [{\citenamefont {Arvanitaki}\ \emph {et~al.}(2015)\citenamefont
  {Arvanitaki}, \citenamefont {Baryakhtar},\ and\ \citenamefont
  {Huang}}]{Arvanitaki:2014wva}%
  \BibitemOpen
  \bibfield  {author} {\bibinfo {author} {\bibfnamefont {A.}~\bibnamefont
  {Arvanitaki}}, \bibinfo {author} {\bibfnamefont {M.}~\bibnamefont
  {Baryakhtar}},\ and\ \bibinfo {author} {\bibfnamefont {X.}~\bibnamefont
  {Huang}},\ }\bibfield  {title} {\bibinfo {title} {{Discovering the QCD Axion
  with Black Holes and Gravitational Waves}},\ }\href
  {https://doi.org/10.1103/PhysRevD.91.084011} {\bibfield  {journal} {\bibinfo
  {journal} {Phys. Rev. D}\ }\textbf {\bibinfo {volume} {91}},\ \bibinfo
  {pages} {084011} (\bibinfo {year} {2015})},\ \Eprint
  {https://arxiv.org/abs/1411.2263} {arXiv:1411.2263 [hep-ph]} \BibitemShut
  {NoStop}%
\bibitem [{\citenamefont {Herdeiro}\ and\ \citenamefont
  {Radu}(2014)}]{Herdeiro:2014goa}%
  \BibitemOpen
  \bibfield  {author} {\bibinfo {author} {\bibfnamefont {C.~A.~R.}\
  \bibnamefont {Herdeiro}}\ and\ \bibinfo {author} {\bibfnamefont
  {E.}~\bibnamefont {Radu}},\ }\bibfield  {title} {\bibinfo {title} {{Kerr
  black holes with scalar hair}},\ }\href
  {https://doi.org/10.1103/PhysRevLett.112.221101} {\bibfield  {journal}
  {\bibinfo  {journal} {Phys. Rev. Lett.}\ }\textbf {\bibinfo {volume} {112}},\
  \bibinfo {pages} {221101} (\bibinfo {year} {2014})},\ \Eprint
  {https://arxiv.org/abs/1403.2757} {arXiv:1403.2757 [gr-qc]} \BibitemShut
  {NoStop}%
\bibitem [{\citenamefont {East}\ and\ \citenamefont
  {Pretorius}(2017)}]{East:2017ovw}%
  \BibitemOpen
  \bibfield  {author} {\bibinfo {author} {\bibfnamefont {W.~E.}\ \bibnamefont
  {East}}\ and\ \bibinfo {author} {\bibfnamefont {F.}~\bibnamefont
  {Pretorius}},\ }\bibfield  {title} {\bibinfo {title} {{Superradiant
  Instability and Backreaction of Massive Vector Fields around Kerr Black
  Holes}},\ }\href {https://doi.org/10.1103/PhysRevLett.119.041101} {\bibfield
  {journal} {\bibinfo  {journal} {Phys. Rev. Lett.}\ }\textbf {\bibinfo
  {volume} {119}},\ \bibinfo {pages} {041101} (\bibinfo {year} {2017})},\
  \Eprint {https://arxiv.org/abs/1704.04791} {arXiv:1704.04791 [gr-qc]}
  \BibitemShut {NoStop}%
\bibitem [{\citenamefont {Brito}\ \emph
  {et~al.}(2015{\natexlab{a}})\citenamefont {Brito}, \citenamefont {Cardoso},\
  and\ \citenamefont {Pani}}]{Brito:2015oca}%
  \BibitemOpen
  \bibfield  {author} {\bibinfo {author} {\bibfnamefont {R.}~\bibnamefont
  {Brito}}, \bibinfo {author} {\bibfnamefont {V.}~\bibnamefont {Cardoso}},\
  and\ \bibinfo {author} {\bibfnamefont {P.}~\bibnamefont {Pani}},\ }\bibfield
  {title} {\bibinfo {title} {{Superradiance}: {New Frontiers in Black Hole
  Physics}},\ }\href {https://doi.org/10.1007/978-3-319-19000-6} {\bibfield
  {journal} {\bibinfo  {journal} {Lect. Notes Phys.}\ }\textbf {\bibinfo
  {volume} {906}},\ \bibinfo {pages} {pp.1} (\bibinfo {year}
  {2015}{\natexlab{a}})},\ \Eprint {https://arxiv.org/abs/1501.06570}
  {arXiv:1501.06570 [gr-qc]} \BibitemShut {NoStop}%
\bibitem [{\citenamefont {Gondolo}\ and\ \citenamefont
  {Silk}(1999)}]{Gondolo:1999ef}%
  \BibitemOpen
  \bibfield  {author} {\bibinfo {author} {\bibfnamefont {P.}~\bibnamefont
  {Gondolo}}\ and\ \bibinfo {author} {\bibfnamefont {J.}~\bibnamefont {Silk}},\
  }\bibfield  {title} {\bibinfo {title} {{Dark matter annihilation at the
  galactic center}},\ }\href {https://doi.org/10.1103/PhysRevLett.83.1719}
  {\bibfield  {journal} {\bibinfo  {journal} {Phys. Rev. Lett.}\ }\textbf
  {\bibinfo {volume} {83}},\ \bibinfo {pages} {1719} (\bibinfo {year}
  {1999})},\ \Eprint {https://arxiv.org/abs/astro-ph/9906391}
  {arXiv:astro-ph/9906391} \BibitemShut {NoStop}%
\bibitem [{\citenamefont {Sadeghian}\ \emph {et~al.}(2013)\citenamefont
  {Sadeghian}, \citenamefont {Ferrer},\ and\ \citenamefont
  {Will}}]{Sadeghian:2013laa}%
  \BibitemOpen
  \bibfield  {author} {\bibinfo {author} {\bibfnamefont {L.}~\bibnamefont
  {Sadeghian}}, \bibinfo {author} {\bibfnamefont {F.}~\bibnamefont {Ferrer}},\
  and\ \bibinfo {author} {\bibfnamefont {C.~M.}\ \bibnamefont {Will}},\
  }\bibfield  {title} {\bibinfo {title} {{Dark matter distributions around
  massive black holes: A general relativistic analysis}},\ }\href
  {https://doi.org/10.1103/PhysRevD.88.063522} {\bibfield  {journal} {\bibinfo
  {journal} {Phys. Rev. D}\ }\textbf {\bibinfo {volume} {88}},\ \bibinfo
  {pages} {063522} (\bibinfo {year} {2013})},\ \Eprint
  {https://arxiv.org/abs/1305.2619} {arXiv:1305.2619 [astro-ph.GA]}
  \BibitemShut {NoStop}%
\bibitem [{\citenamefont {Ferrer}\ \emph {et~al.}(2017)\citenamefont {Ferrer},
  \citenamefont {da~Rosa},\ and\ \citenamefont {Will}}]{Ferrer:2017xwm}%
  \BibitemOpen
  \bibfield  {author} {\bibinfo {author} {\bibfnamefont {F.}~\bibnamefont
  {Ferrer}}, \bibinfo {author} {\bibfnamefont {A.~M.}\ \bibnamefont
  {da~Rosa}},\ and\ \bibinfo {author} {\bibfnamefont {C.~M.}\ \bibnamefont
  {Will}},\ }\bibfield  {title} {\bibinfo {title} {{Dark matter spikes in the
  vicinity of Kerr black holes}},\ }\href
  {https://doi.org/10.1103/PhysRevD.96.083014} {\bibfield  {journal} {\bibinfo
  {journal} {Phys. Rev. D}\ }\textbf {\bibinfo {volume} {96}},\ \bibinfo
  {pages} {083014} (\bibinfo {year} {2017})},\ \Eprint
  {https://arxiv.org/abs/1707.06302} {arXiv:1707.06302 [astro-ph.CO]}
  \BibitemShut {NoStop}%
\bibitem [{\citenamefont {Speeney}\ \emph {et~al.}(2022)\citenamefont
  {Speeney}, \citenamefont {Antonelli}, \citenamefont {Baibhav},\ and\
  \citenamefont {Berti}}]{Speeney:2022ryg}%
  \BibitemOpen
  \bibfield  {author} {\bibinfo {author} {\bibfnamefont {N.}~\bibnamefont
  {Speeney}}, \bibinfo {author} {\bibfnamefont {A.}~\bibnamefont {Antonelli}},
  \bibinfo {author} {\bibfnamefont {V.}~\bibnamefont {Baibhav}},\ and\ \bibinfo
  {author} {\bibfnamefont {E.}~\bibnamefont {Berti}},\ }\bibfield  {title}
  {\bibinfo {title} {{Impact of relativistic corrections on the detectability
  of dark-matter spikes with gravitational waves}},\ }\href
  {https://doi.org/10.1103/PhysRevD.106.044027} {\bibfield  {journal} {\bibinfo
   {journal} {Phys. Rev. D}\ }\textbf {\bibinfo {volume} {106}},\ \bibinfo
  {pages} {044027} (\bibinfo {year} {2022})},\ \Eprint
  {https://arxiv.org/abs/2204.12508} {arXiv:2204.12508 [gr-qc]} \BibitemShut
  {NoStop}%
\bibitem [{\citenamefont {Speeney}\ \emph {et~al.}(2024)\citenamefont
  {Speeney}, \citenamefont {Berti}, \citenamefont {Cardoso},\ and\
  \citenamefont {Maselli}}]{Speeney:2024mas}%
  \BibitemOpen
  \bibfield  {author} {\bibinfo {author} {\bibfnamefont {N.}~\bibnamefont
  {Speeney}}, \bibinfo {author} {\bibfnamefont {E.}~\bibnamefont {Berti}},
  \bibinfo {author} {\bibfnamefont {V.}~\bibnamefont {Cardoso}},\ and\ \bibinfo
  {author} {\bibfnamefont {A.}~\bibnamefont {Maselli}},\ }\bibfield  {title}
  {\bibinfo {title} {{Black holes surrounded by generic matter distributions:
  polar perturbations and energy flux}},\ }\href@noop {} {\  (\bibinfo {year}
  {2024})},\ \Eprint {https://arxiv.org/abs/2401.00932} {arXiv:2401.00932
  [gr-qc]} \BibitemShut {NoStop}%
\bibitem [{\citenamefont {Gnedin}\ and\ \citenamefont
  {Primack}(2004)}]{Gnedin:2003rj}%
  \BibitemOpen
  \bibfield  {author} {\bibinfo {author} {\bibfnamefont {O.~Y.}\ \bibnamefont
  {Gnedin}}\ and\ \bibinfo {author} {\bibfnamefont {J.~R.}\ \bibnamefont
  {Primack}},\ }\bibfield  {title} {\bibinfo {title} {{Dark Matter Profile in
  the Galactic Center}},\ }\href
  {https://doi.org/10.1103/PhysRevLett.93.061302} {\bibfield  {journal}
  {\bibinfo  {journal} {Phys. Rev. Lett.}\ }\textbf {\bibinfo {volume} {93}},\
  \bibinfo {pages} {061302} (\bibinfo {year} {2004})},\ \Eprint
  {https://arxiv.org/abs/astro-ph/0308385} {arXiv:astro-ph/0308385}
  \BibitemShut {NoStop}%
\bibitem [{\citenamefont {Merritt}(2004)}]{Merritt:2003qk}%
  \BibitemOpen
  \bibfield  {author} {\bibinfo {author} {\bibfnamefont {D.}~\bibnamefont
  {Merritt}},\ }\bibfield  {title} {\bibinfo {title} {{Evolution of the dark
  matter distribution at the galactic center}},\ }\href
  {https://doi.org/10.1103/PhysRevLett.92.201304} {\bibfield  {journal}
  {\bibinfo  {journal} {Phys. Rev. Lett.}\ }\textbf {\bibinfo {volume} {92}},\
  \bibinfo {pages} {201304} (\bibinfo {year} {2004})},\ \Eprint
  {https://arxiv.org/abs/astro-ph/0311594} {arXiv:astro-ph/0311594}
  \BibitemShut {NoStop}%
\bibitem [{\citenamefont {Merritt}\ \emph {et~al.}(2007)\citenamefont
  {Merritt}, \citenamefont {Harfst},\ and\ \citenamefont
  {Bertone}}]{Merritt:2006mt}%
  \BibitemOpen
  \bibfield  {author} {\bibinfo {author} {\bibfnamefont {D.}~\bibnamefont
  {Merritt}}, \bibinfo {author} {\bibfnamefont {S.}~\bibnamefont {Harfst}},\
  and\ \bibinfo {author} {\bibfnamefont {G.}~\bibnamefont {Bertone}},\
  }\bibfield  {title} {\bibinfo {title} {{Collisionally Regenerated Dark Matter
  Structures in Galactic Nuclei}},\ }\href
  {https://doi.org/10.1103/PhysRevD.75.043517} {\bibfield  {journal} {\bibinfo
  {journal} {Phys. Rev. D}\ }\textbf {\bibinfo {volume} {75}},\ \bibinfo
  {pages} {043517} (\bibinfo {year} {2007})},\ \Eprint
  {https://arxiv.org/abs/astro-ph/0610425} {arXiv:astro-ph/0610425}
  \BibitemShut {NoStop}%
\bibitem [{\citenamefont {Shapiro}\ and\ \citenamefont
  {Heggie}(2022)}]{Shapiro:2022prq}%
  \BibitemOpen
  \bibfield  {author} {\bibinfo {author} {\bibfnamefont {S.~L.}\ \bibnamefont
  {Shapiro}}\ and\ \bibinfo {author} {\bibfnamefont {D.~C.}\ \bibnamefont
  {Heggie}},\ }\bibfield  {title} {\bibinfo {title} {{Effect of stars on the
  dark matter spike around a black hole: A tale of two treatments}},\ }\href
  {https://doi.org/10.1103/PhysRevD.106.043018} {\bibfield  {journal} {\bibinfo
   {journal} {Phys. Rev. D}\ }\textbf {\bibinfo {volume} {106}},\ \bibinfo
  {pages} {043018} (\bibinfo {year} {2022})},\ \Eprint
  {https://arxiv.org/abs/2209.08105} {arXiv:2209.08105 [astro-ph.GA]}
  \BibitemShut {NoStop}%
\bibitem [{\citenamefont {De~Luca}\ and\ \citenamefont
  {Khoury}(2023)}]{DeLuca:2023laa}%
  \BibitemOpen
  \bibfield  {author} {\bibinfo {author} {\bibfnamefont {V.}~\bibnamefont
  {De~Luca}}\ and\ \bibinfo {author} {\bibfnamefont {J.}~\bibnamefont
  {Khoury}},\ }\bibfield  {title} {\bibinfo {title} {{Superfluid dark matter
  around black holes}},\ }\href {https://doi.org/10.1088/1475-7516/2023/04/048}
  {\bibfield  {journal} {\bibinfo  {journal} {JCAP}\ }\textbf {\bibinfo
  {volume} {04}},\ \bibinfo {pages} {048}},\ \Eprint
  {https://arxiv.org/abs/2302.10286} {arXiv:2302.10286 [astro-ph.CO]}
  \BibitemShut {NoStop}%
\bibitem [{\citenamefont {Berezhiani}\ \emph {et~al.}(2023)\citenamefont
  {Berezhiani}, \citenamefont {Cintia}, \citenamefont {De~Luca},\ and\
  \citenamefont {Khoury}}]{Berezhiani:2023vlo}%
  \BibitemOpen
  \bibfield  {author} {\bibinfo {author} {\bibfnamefont {L.}~\bibnamefont
  {Berezhiani}}, \bibinfo {author} {\bibfnamefont {G.}~\bibnamefont {Cintia}},
  \bibinfo {author} {\bibfnamefont {V.}~\bibnamefont {De~Luca}},\ and\ \bibinfo
  {author} {\bibfnamefont {J.}~\bibnamefont {Khoury}},\ }\bibfield  {title}
  {\bibinfo {title} {{Dynamical friction in dark matter superfluids: The
  evolution of black hole binaries}},\ }\href@noop {} {\  (\bibinfo {year}
  {2023})},\ \Eprint {https://arxiv.org/abs/2311.07672} {arXiv:2311.07672
  [astro-ph.CO]} \BibitemShut {NoStop}%
\bibitem [{\citenamefont {Sanchis-Gual}\ \emph {et~al.}(2016)\citenamefont
  {Sanchis-Gual}, \citenamefont {Degollado}, \citenamefont {Izquierdo},
  \citenamefont {Font},\ and\ \citenamefont {Montero}}]{Sanchis-Gual:2016jst}%
  \BibitemOpen
  \bibfield  {author} {\bibinfo {author} {\bibfnamefont {N.}~\bibnamefont
  {Sanchis-Gual}}, \bibinfo {author} {\bibfnamefont {J.~C.}\ \bibnamefont
  {Degollado}}, \bibinfo {author} {\bibfnamefont {P.}~\bibnamefont
  {Izquierdo}}, \bibinfo {author} {\bibfnamefont {J.~A.}\ \bibnamefont
  {Font}},\ and\ \bibinfo {author} {\bibfnamefont {P.~J.}\ \bibnamefont
  {Montero}},\ }\bibfield  {title} {\bibinfo {title} {{Quasistationary
  solutions of scalar fields around accreting black holes}},\ }\href
  {https://doi.org/10.1103/PhysRevD.94.043004} {\bibfield  {journal} {\bibinfo
  {journal} {Phys. Rev. D}\ }\textbf {\bibinfo {volume} {94}},\ \bibinfo
  {pages} {043004} (\bibinfo {year} {2016})},\ \Eprint
  {https://arxiv.org/abs/1606.05146} {arXiv:1606.05146 [gr-qc]} \BibitemShut
  {NoStop}%
\bibitem [{\citenamefont {Clough}\ \emph {et~al.}(2019)\citenamefont {Clough},
  \citenamefont {Ferreira},\ and\ \citenamefont {Lagos}}]{Clough:2019jpm}%
  \BibitemOpen
  \bibfield  {author} {\bibinfo {author} {\bibfnamefont {K.}~\bibnamefont
  {Clough}}, \bibinfo {author} {\bibfnamefont {P.~G.}\ \bibnamefont
  {Ferreira}},\ and\ \bibinfo {author} {\bibfnamefont {M.}~\bibnamefont
  {Lagos}},\ }\bibfield  {title} {\bibinfo {title} {{Growth of massive scalar
  hair around a Schwarzschild black hole}},\ }\href
  {https://doi.org/10.1103/PhysRevD.100.063014} {\bibfield  {journal} {\bibinfo
   {journal} {Phys. Rev. D}\ }\textbf {\bibinfo {volume} {100}},\ \bibinfo
  {pages} {063014} (\bibinfo {year} {2019})},\ \Eprint
  {https://arxiv.org/abs/1904.12783} {arXiv:1904.12783 [gr-qc]} \BibitemShut
  {NoStop}%
\bibitem [{\citenamefont {Hui}\ \emph {et~al.}(2019)\citenamefont {Hui},
  \citenamefont {Kabat}, \citenamefont {Li}, \citenamefont {Santoni},\ and\
  \citenamefont {Wong}}]{Hui:2019aqm}%
  \BibitemOpen
  \bibfield  {author} {\bibinfo {author} {\bibfnamefont {L.}~\bibnamefont
  {Hui}}, \bibinfo {author} {\bibfnamefont {D.}~\bibnamefont {Kabat}}, \bibinfo
  {author} {\bibfnamefont {X.}~\bibnamefont {Li}}, \bibinfo {author}
  {\bibfnamefont {L.}~\bibnamefont {Santoni}},\ and\ \bibinfo {author}
  {\bibfnamefont {S.~S.~C.}\ \bibnamefont {Wong}},\ }\bibfield  {title}
  {\bibinfo {title} {{Black Hole Hair from Scalar Dark Matter}},\ }\href
  {https://doi.org/10.1088/1475-7516/2019/06/038} {\bibfield  {journal}
  {\bibinfo  {journal} {JCAP}\ }\textbf {\bibinfo {volume} {06}},\ \bibinfo
  {pages} {038}},\ \Eprint {https://arxiv.org/abs/1904.12803} {arXiv:1904.12803
  [gr-qc]} \BibitemShut {NoStop}%
\bibitem [{\citenamefont {Bamber}\ \emph {et~al.}(2021)\citenamefont {Bamber},
  \citenamefont {Clough}, \citenamefont {Ferreira}, \citenamefont {Hui},\ and\
  \citenamefont {Lagos}}]{Bamber:2020bpu}%
  \BibitemOpen
  \bibfield  {author} {\bibinfo {author} {\bibfnamefont {J.}~\bibnamefont
  {Bamber}}, \bibinfo {author} {\bibfnamefont {K.}~\bibnamefont {Clough}},
  \bibinfo {author} {\bibfnamefont {P.~G.}\ \bibnamefont {Ferreira}}, \bibinfo
  {author} {\bibfnamefont {L.}~\bibnamefont {Hui}},\ and\ \bibinfo {author}
  {\bibfnamefont {M.}~\bibnamefont {Lagos}},\ }\bibfield  {title} {\bibinfo
  {title} {{Growth of accretion driven scalar hair around Kerr black holes}},\
  }\href {https://doi.org/10.1103/PhysRevD.103.044059} {\bibfield  {journal}
  {\bibinfo  {journal} {Phys. Rev. D}\ }\textbf {\bibinfo {volume} {103}},\
  \bibinfo {pages} {044059} (\bibinfo {year} {2021})},\ \Eprint
  {https://arxiv.org/abs/2011.07870} {arXiv:2011.07870 [gr-qc]} \BibitemShut
  {NoStop}%
\bibitem [{\citenamefont {Bucciotti}\ and\ \citenamefont
  {Trincherini}(2023)}]{Bucciotti:2023bvw}%
  \BibitemOpen
  \bibfield  {author} {\bibinfo {author} {\bibfnamefont {B.}~\bibnamefont
  {Bucciotti}}\ and\ \bibinfo {author} {\bibfnamefont {E.}~\bibnamefont
  {Trincherini}},\ }\bibfield  {title} {\bibinfo {title} {{interplay between
  black holes and ultralight dark matter: analytic solutions}},\ }\href
  {https://doi.org/10.1007/JHEP11(2023)193} {\bibfield  {journal} {\bibinfo
  {journal} {JHEP}\ }\textbf {\bibinfo {volume} {11}},\ \bibinfo {pages}
  {193}},\ \Eprint {https://arxiv.org/abs/2309.02482} {arXiv:2309.02482
  [hep-th]} \BibitemShut {NoStop}%
\bibitem [{\citenamefont {de~Cesare}\ and\ \citenamefont
  {Oliveri}(2023)}]{deCesare:2023rmg}%
  \BibitemOpen
  \bibfield  {author} {\bibinfo {author} {\bibfnamefont {M.}~\bibnamefont
  {de~Cesare}}\ and\ \bibinfo {author} {\bibfnamefont {R.}~\bibnamefont
  {Oliveri}},\ }\bibfield  {title} {\bibinfo {title} {{Backreaction of scalar
  waves on black holes at low frequencies}},\ }\href
  {https://doi.org/10.1103/PhysRevD.108.044050} {\bibfield  {journal} {\bibinfo
   {journal} {Phys. Rev. D}\ }\textbf {\bibinfo {volume} {108}},\ \bibinfo
  {pages} {044050} (\bibinfo {year} {2023})},\ \Eprint
  {https://arxiv.org/abs/2305.04970} {arXiv:2305.04970 [gr-qc]} \BibitemShut
  {NoStop}%
\bibitem [{\citenamefont {Bamber}\ \emph {et~al.}(2023)\citenamefont {Bamber},
  \citenamefont {Aurrekoetxea}, \citenamefont {Clough},\ and\ \citenamefont
  {Ferreira}}]{Bamber:2022pbs}%
  \BibitemOpen
  \bibfield  {author} {\bibinfo {author} {\bibfnamefont {J.}~\bibnamefont
  {Bamber}}, \bibinfo {author} {\bibfnamefont {J.~C.}\ \bibnamefont
  {Aurrekoetxea}}, \bibinfo {author} {\bibfnamefont {K.}~\bibnamefont
  {Clough}},\ and\ \bibinfo {author} {\bibfnamefont {P.~G.}\ \bibnamefont
  {Ferreira}},\ }\bibfield  {title} {\bibinfo {title} {{Black hole merger
  simulations in wave dark matter environments}},\ }\href
  {https://doi.org/10.1103/PhysRevD.107.024035} {\bibfield  {journal} {\bibinfo
   {journal} {Phys. Rev. D}\ }\textbf {\bibinfo {volume} {107}},\ \bibinfo
  {pages} {024035} (\bibinfo {year} {2023})},\ \Eprint
  {https://arxiv.org/abs/2210.09254} {arXiv:2210.09254 [gr-qc]} \BibitemShut
  {NoStop}%
\bibitem [{\citenamefont {Aurrekoetxea}\ \emph
  {et~al.}(2023{\natexlab{a}})\citenamefont {Aurrekoetxea}, \citenamefont
  {Clough}, \citenamefont {Bamber},\ and\ \citenamefont
  {Ferreira}}]{Aurrekoetxea:2023jwk}%
  \BibitemOpen
  \bibfield  {author} {\bibinfo {author} {\bibfnamefont {J.~C.}\ \bibnamefont
  {Aurrekoetxea}}, \bibinfo {author} {\bibfnamefont {K.}~\bibnamefont
  {Clough}}, \bibinfo {author} {\bibfnamefont {J.}~\bibnamefont {Bamber}},\
  and\ \bibinfo {author} {\bibfnamefont {P.~G.}\ \bibnamefont {Ferreira}},\
  }\bibfield  {title} {\bibinfo {title} {{The effect of wave dark matter on
  equal mass black hole mergers}},\ }\href@noop {} {\  (\bibinfo {year}
  {2023}{\natexlab{a}})},\ \Eprint {https://arxiv.org/abs/2311.18156}
  {arXiv:2311.18156 [gr-qc]} \BibitemShut {NoStop}%
\bibitem [{\citenamefont {Marsh}(2016)}]{Marsh:2015xka}%
  \BibitemOpen
  \bibfield  {author} {\bibinfo {author} {\bibfnamefont {D.~J.~E.}\
  \bibnamefont {Marsh}},\ }\bibfield  {title} {\bibinfo {title} {{Axion
  Cosmology}},\ }\href {https://doi.org/10.1016/j.physrep.2016.06.005}
  {\bibfield  {journal} {\bibinfo  {journal} {Phys. Rept.}\ }\textbf {\bibinfo
  {volume} {643}},\ \bibinfo {pages} {1} (\bibinfo {year} {2016})},\ \Eprint
  {https://arxiv.org/abs/1510.07633} {arXiv:1510.07633 [astro-ph.CO]}
  \BibitemShut {NoStop}%
\bibitem [{\citenamefont {Peccei}(2008)}]{Peccei:2006as}%
  \BibitemOpen
  \bibfield  {author} {\bibinfo {author} {\bibfnamefont {R.~D.}\ \bibnamefont
  {Peccei}},\ }\bibfield  {title} {\bibinfo {title} {{The Strong CP problem and
  axions}},\ }\href {https://doi.org/10.1007/978-3-540-73518-2_1} {\bibfield
  {journal} {\bibinfo  {journal} {Lect. Notes Phys.}\ }\textbf {\bibinfo
  {volume} {741}},\ \bibinfo {pages} {3} (\bibinfo {year} {2008})},\ \Eprint
  {https://arxiv.org/abs/hep-ph/0607268} {arXiv:hep-ph/0607268} \BibitemShut
  {NoStop}%
\bibitem [{\citenamefont {Aurrekoetxea}\ \emph
  {et~al.}(2023{\natexlab{b}})\citenamefont {Aurrekoetxea}, \citenamefont
  {Bamber}, \citenamefont {Brady}, \citenamefont {Clough}, \citenamefont
  {Helfer}, \citenamefont {Marsden}, \citenamefont {Traykova},\ and\
  \citenamefont {Wang}}]{Aurrekoetxea:2023fhl}%
  \BibitemOpen
  \bibfield  {author} {\bibinfo {author} {\bibfnamefont {J.~C.}\ \bibnamefont
  {Aurrekoetxea}}, \bibinfo {author} {\bibfnamefont {J.}~\bibnamefont
  {Bamber}}, \bibinfo {author} {\bibfnamefont {S.~E.}\ \bibnamefont {Brady}},
  \bibinfo {author} {\bibfnamefont {K.}~\bibnamefont {Clough}}, \bibinfo
  {author} {\bibfnamefont {T.}~\bibnamefont {Helfer}}, \bibinfo {author}
  {\bibfnamefont {J.}~\bibnamefont {Marsden}}, \bibinfo {author} {\bibfnamefont
  {D.}~\bibnamefont {Traykova}},\ and\ \bibinfo {author} {\bibfnamefont
  {Z.}~\bibnamefont {Wang}},\ }\bibfield  {title} {\bibinfo {title}
  {{GRDzhadzha: A code for evolving relativistic matter on analytic metric
  backgrounds}},\ }\href@noop {} {\  (\bibinfo {year} {2023}{\natexlab{b}})},\
  \Eprint {https://arxiv.org/abs/2308.08299} {arXiv:2308.08299 [gr-qc]}
  \BibitemShut {NoStop}%
\bibitem [{\citenamefont {Okawa}\ \emph {et~al.}(2014)\citenamefont {Okawa},
  \citenamefont {Witek},\ and\ \citenamefont {Cardoso}}]{Okawa:2014nda}%
  \BibitemOpen
  \bibfield  {author} {\bibinfo {author} {\bibfnamefont {H.}~\bibnamefont
  {Okawa}}, \bibinfo {author} {\bibfnamefont {H.}~\bibnamefont {Witek}},\ and\
  \bibinfo {author} {\bibfnamefont {V.}~\bibnamefont {Cardoso}},\ }\bibfield
  {title} {\bibinfo {title} {{Black holes and fundamental fields in Numerical
  Relativity: initial data construction and evolution of bound states}},\
  }\href {https://doi.org/10.1103/PhysRevD.89.104032} {\bibfield  {journal}
  {\bibinfo  {journal} {Phys. Rev. D}\ }\textbf {\bibinfo {volume} {89}},\
  \bibinfo {pages} {104032} (\bibinfo {year} {2014})},\ \Eprint
  {https://arxiv.org/abs/1401.1548} {arXiv:1401.1548 [gr-qc]} \BibitemShut
  {NoStop}%
\bibitem [{\citenamefont {Clough}(2021)}]{Clough:2021qlv}%
  \BibitemOpen
  \bibfield  {author} {\bibinfo {author} {\bibfnamefont {K.}~\bibnamefont
  {Clough}},\ }\bibfield  {title} {\bibinfo {title} {{Continuity equations for
  general matter: applications in numerical relativity}},\ }\href
  {https://doi.org/10.1088/1361-6382/ac10ee} {\bibfield  {journal} {\bibinfo
  {journal} {Class. Quant. Grav.}\ }\textbf {\bibinfo {volume} {38}},\ \bibinfo
  {pages} {167001} (\bibinfo {year} {2021})},\ \Eprint
  {https://arxiv.org/abs/2104.13420} {arXiv:2104.13420 [gr-qc]} \BibitemShut
  {NoStop}%
\bibitem [{\citenamefont {Brito}\ \emph
  {et~al.}(2015{\natexlab{b}})\citenamefont {Brito}, \citenamefont {Cardoso},\
  and\ \citenamefont {Pani}}]{Brito:2014wla}%
  \BibitemOpen
  \bibfield  {author} {\bibinfo {author} {\bibfnamefont {R.}~\bibnamefont
  {Brito}}, \bibinfo {author} {\bibfnamefont {V.}~\bibnamefont {Cardoso}},\
  and\ \bibinfo {author} {\bibfnamefont {P.}~\bibnamefont {Pani}},\ }\bibfield
  {title} {\bibinfo {title} {{Black holes as particle detectors: evolution of
  superradiant instabilities}},\ }\href
  {https://doi.org/10.1088/0264-9381/32/13/134001} {\bibfield  {journal}
  {\bibinfo  {journal} {Class. Quant. Grav.}\ }\textbf {\bibinfo {volume}
  {32}},\ \bibinfo {pages} {134001} (\bibinfo {year} {2015}{\natexlab{b}})},\
  \Eprint {https://arxiv.org/abs/1411.0686} {arXiv:1411.0686 [gr-qc]}
  \BibitemShut {NoStop}%
\bibitem [{\citenamefont {East}(2018)}]{East:2018glu}%
  \BibitemOpen
  \bibfield  {author} {\bibinfo {author} {\bibfnamefont {W.~E.}\ \bibnamefont
  {East}},\ }\bibfield  {title} {\bibinfo {title} {{Massive Boson Superradiant
  Instability of Black Holes: Nonlinear Growth, Saturation, and Gravitational
  Radiation}},\ }\href {https://doi.org/10.1103/PhysRevLett.121.131104}
  {\bibfield  {journal} {\bibinfo  {journal} {Phys. Rev. Lett.}\ }\textbf
  {\bibinfo {volume} {121}},\ \bibinfo {pages} {131104} (\bibinfo {year}
  {2018})},\ \Eprint {https://arxiv.org/abs/1807.00043} {arXiv:1807.00043
  [gr-qc]} \BibitemShut {NoStop}%
\bibitem [{\citenamefont {Clough}\ \emph {et~al.}(2015)\citenamefont {Clough},
  \citenamefont {Figueras}, \citenamefont {Finkel}, \citenamefont {Kunesch},
  \citenamefont {Lim},\ and\ \citenamefont {Tunyasuvunakool}}]{Clough:2015sqa}%
  \BibitemOpen
  \bibfield  {author} {\bibinfo {author} {\bibfnamefont {K.}~\bibnamefont
  {Clough}}, \bibinfo {author} {\bibfnamefont {P.}~\bibnamefont {Figueras}},
  \bibinfo {author} {\bibfnamefont {H.}~\bibnamefont {Finkel}}, \bibinfo
  {author} {\bibfnamefont {M.}~\bibnamefont {Kunesch}}, \bibinfo {author}
  {\bibfnamefont {E.~A.}\ \bibnamefont {Lim}},\ and\ \bibinfo {author}
  {\bibfnamefont {S.}~\bibnamefont {Tunyasuvunakool}},\ }\bibfield  {title}
  {\bibinfo {title} {{GRChombo : Numerical Relativity with Adaptive Mesh
  Refinement}},\ }\href {https://doi.org/10.1088/0264-9381/32/24/245011}
  {\bibfield  {journal} {\bibinfo  {journal} {Class. Quant. Grav.}\ }\textbf
  {\bibinfo {volume} {32}},\ \bibinfo {pages} {245011} (\bibinfo {year}
  {2015})},\ \Eprint {https://arxiv.org/abs/1503.03436} {arXiv:1503.03436
  [gr-qc]} \BibitemShut {NoStop}%
\bibitem [{\citenamefont {Andrade}\ \emph {et~al.}(2021)\citenamefont {Andrade}
  \emph {et~al.}}]{Andrade:2021rbd}%
  \BibitemOpen
  \bibfield  {author} {\bibinfo {author} {\bibfnamefont {T.}~\bibnamefont
  {Andrade}} \emph {et~al.},\ }\bibfield  {title} {\bibinfo {title} {{GRChombo:
  An adaptable numerical relativity code for fundamental physics}},\ }\href
  {https://doi.org/10.21105/joss.03703} {\bibfield  {journal} {\bibinfo
  {journal} {J. Open Source Softw.}\ }\textbf {\bibinfo {volume} {6}},\
  \bibinfo {pages} {3703} (\bibinfo {year} {2021})},\ \Eprint
  {https://arxiv.org/abs/2201.03458} {arXiv:2201.03458 [gr-qc]} \BibitemShut
  {NoStop}%
\bibitem [{\citenamefont {Dyson}\ \emph {et~al.}(2024)\citenamefont {Dyson},
  \citenamefont {Redondo-Yuste}, \citenamefont {van~de Meent},\ and\
  \citenamefont {Cardoso}}]{Dyson:2024qrq}%
  \BibitemOpen
  \bibfield  {author} {\bibinfo {author} {\bibfnamefont {C.}~\bibnamefont
  {Dyson}}, \bibinfo {author} {\bibfnamefont {J.}~\bibnamefont
  {Redondo-Yuste}}, \bibinfo {author} {\bibfnamefont {M.}~\bibnamefont {van~de
  Meent}},\ and\ \bibinfo {author} {\bibfnamefont {V.}~\bibnamefont
  {Cardoso}},\ }\bibfield  {title} {\bibinfo {title} {{Relativistic
  aerodynamics of spinning black holes}},\ }\href@noop {} {\  (\bibinfo {year}
  {2024})},\ \Eprint {https://arxiv.org/abs/2402.07981} {arXiv:2402.07981
  [gr-qc]} \BibitemShut {NoStop}%
\bibitem [{\citenamefont {Stanzione}\ \emph {et~al.}(2020)\citenamefont
  {Stanzione}, \citenamefont {West}, \citenamefont {Evans}, \citenamefont
  {Minyard}, \citenamefont {Ghattas},\ and\ \citenamefont
  {Panda}}]{10.1145/3311790.3396656}%
  \BibitemOpen
  \bibfield  {author} {\bibinfo {author} {\bibfnamefont {D.}~\bibnamefont
  {Stanzione}}, \bibinfo {author} {\bibfnamefont {J.}~\bibnamefont {West}},
  \bibinfo {author} {\bibfnamefont {R.~T.}\ \bibnamefont {Evans}}, \bibinfo
  {author} {\bibfnamefont {T.}~\bibnamefont {Minyard}}, \bibinfo {author}
  {\bibfnamefont {O.}~\bibnamefont {Ghattas}},\ and\ \bibinfo {author}
  {\bibfnamefont {D.~K.}\ \bibnamefont {Panda}},\ }\bibfield  {title} {\bibinfo
  {title} {Frontera: The evolution of leadership computing at the national
  science foundation},\ }in\ \href {https://doi.org/10.1145/3311790.3396656}
  {\emph {\bibinfo {booktitle} {Practice and Experience in Advanced Research
  Computing}}},\ \bibinfo {series and number} {PEARC ’20}\ (\bibinfo
  {publisher} {Association for Computing Machinery},\ \bibinfo {address} {New
  York, NY, USA},\ \bibinfo {year} {2020})\ p.\ \bibinfo {pages}
  {106–111}\BibitemShut {NoStop}%
\bibitem [{\citenamefont {Boyer}\ and\ \citenamefont
  {Lindquist}(1967)}]{Boyer:1966qh}%
  \BibitemOpen
  \bibfield  {author} {\bibinfo {author} {\bibfnamefont {R.~H.}\ \bibnamefont
  {Boyer}}\ and\ \bibinfo {author} {\bibfnamefont {R.~W.}\ \bibnamefont
  {Lindquist}},\ }\bibfield  {title} {\bibinfo {title} {{Maximal analytic
  extension of the Kerr metric}},\ }\href {https://doi.org/10.1063/1.1705193}
  {\bibfield  {journal} {\bibinfo  {journal} {J. Math. Phys.}\ }\textbf
  {\bibinfo {volume} {8}},\ \bibinfo {pages} {265} (\bibinfo {year}
  {1967})}\BibitemShut {NoStop}%
\end{thebibliography}%

\end{document}